\documentclass[%
twocolumn,
 amsmath,amssymb,
prb,
floatfix,
]{revtex4-2}
\pdfoutput=1

\usepackage{graphicx}% Include figure files
\usepackage{dcolumn}% Align table columns on decimal point
\usepackage{bm}% bold math
\usepackage[normalem]{ulem}
\usepackage{color,soul}

\renewcommand{\thesection}{\Roman{section}} 
\usepackage{float}

\begin{document}

\title[ML-Driven Process of
Alumina Ceramics Laser Machining]{
Machine Learning-Driven Process of
Alumina Ceramics Laser Machining}

\author{Razyeh Behbahani$^{1, 2}$}
\author{Hamidreza Yazdani Sarvestani$^{3}$}
\author{Erfan Fatehi$^{4}$}
\author{Elham Kiyani$^{2,5}$}
\author{Behnam Ashrafi$^{3}$}
\email[]{behnam.ashrafi@cnrc-nrc.gc}
\author{Mikko Karttunen$^{1,2,6}$}
\email[]{mkarttu@uwo.ca}
\author{Meysam Rahmat$^{7}$}

\address{$^1$Department of Physics and Astronomy, The University of Western Ontario, 1151 Richmond Street, London, ON N6A~3K7 Canada}

 \address{$^2$ The Centre for Advanced Materials and Biomaterials (CAMBR), The University of Western Ontario, 1151 Richmond Street, London, ON N6A~5B7 Canada}
 
 \address{$^3$ Aerospace Manufacturing Technology Centre, National Research Council Canada, 5145 Decelles Avenue, Montreal, QC H3T~2B2 Canada}
 
 \address{$^4$ Energy, Mining and Environment, National Research Council Canada, 2620 Speakman Dr. Mississauga, ON L5K~1B1 Canada}
 
 \address{$^5$ Department of Mathematics, Middelesex College, The University of Western Ontario, London, ON N6A~5B7 Canada}
 
 \address{$^6$ Department of Chemistry,The University of Western Ontario, 1151 Richmond Street, London, ON N6A~5B7 Canada}

 \address{$^7$ Aerospace Manufacturing Technology Centre, National Research Council Canada, 1200 Montreal Road, Ottawa, ON K1A~0R6 Canada}
\begin{abstract}
    Laser machining is a highly flexible non-contact manufacturing technique that has been employed widely across academia and industry. Due to nonlinear interactions between light and matter, simulation methods are extremely crucial, as they help enhance the machining quality by offering comprehension of the inter-relationships between the laser processing parameters. On the other hand, experimental processing parameter optimization recommends a systematic, and consequently time-consuming, investigation over the available processing parameter space. An intelligent strategy is to employ machine learning (ML) techniques to capture the relationship between picosecond laser machining parameters for finding proper parameter combinations to create the desired cuts on industrial-grade alumina ceramic with deep, smooth and defect-free patterns. Laser parameters such as beam amplitude and frequency, scanner passing speed and the number of passes over the surface, as well as the vertical distance of the scanner from the sample surface, are used for predicting the depth, top width, and bottom width of the engraved channels using ML models. Owing to the complex correlation between laser parameters, it is shown that Neural Networks (NN) are the most efficient in predicting the outputs. Equipped with an ML model that captures the interconnection between laser parameters and the engraved channel dimensions, one can predict the required input parameters to achieve a target channel geometry. This strategy significantly reduces the cost and effort of experimental laser machining during the development phase, without compromising accuracy or performance. The developed techniques can be applied to a wide range of ceramic laser machining processes.

\end{abstract}

% {\it Keywords:}
%\text
% \end{keywords} 
 %% sn

\maketitle

\section{Introduction}

%\textcolor{red}
{Ceramics are known for their outstanding hardness, thermal performance and corrosion-resistance. These properties make them a suitable candidate for a wide range of applications from automotive industry and space applications to nuclear industry and biomedicine, see for example the review by Greil~\cite{Greil2002-kw}. However, their typically highly oriented crystalline structure results in brittleness and makes them unamenable to a variety of processing techniques and limits their machinability~\cite{Bharathi2021-ty}.
Due to this inherent brittleness, traditional subtractive manufacturing techniques induce microcracks along the cut, which greatly reduce the components’ strengths, especially when subjected to heating cycles~\cite{jianxin2000surface, fenoughty1994machining, li1996surface, boccaccini1997machinability, zavattieri2001grain}.
This has slowed down their adaption in complex parts for applications such as automobile engines, heat exchangers, 
rocket propulsion components, and gas turbines~\cite{rakshit2019review, Bharathi2021-ty}. 
}

%One of the main challenges in processing ceramics is their machinability to produce complex parts for applications such as automobile engines, heat exchangers, high-performance braking systems, space vehicle heat shield systems, rocket propulsion components, and gas turbines~\cite{rakshit2019review, Bharathi2021-ty}. Due to the inherent brittleness of ceramics, traditional subtractive manufacturing techniques induce microcracks along the cut, which greatly reduces the components’ strengths, especially when subjected to heating cycles~\cite{jianxin2000surface, fenoughty1994machining, li1996surface, boccaccini1997machinability, zavattieri2001grain}. 

%\textcolor{red}
{Laser machining is a processing technique, which has attracted attention due to its convenience, efficiency and high precision, as it overcomes many of the problems of subtractive manufacturing~\cite{Cheng2013-ou,bakhtiyari2021review}.
} 
In laser processing, the material is exposed to a high energy beam causing it to rapidly ablate~\cite{ai2020numerical, ai2022investigation} resulting in a superior cut quality and little to no microcrack formation or thermally affected zone~\cite{vsugar2017laser, choudhury2010laser, schulz1993heat}. This capability requires extremely precise and reproducible processing parameter control.

%\textcolor{red}
{Despite of its advantages, laser machining does not come without challenges. One particular one, and the topic of this research, is how to find the optimal laser processing parameters.} 
%\textcolor{red}
{Typically, parameter optimization process must be repeated for any changes in the laser system or material type, which is cumbersome and time consuming. 
To improve and systematize the search,} techniques such as the gradient search, simulation methods and machine learning have been developed to find optimal processing parameters~\cite{mcdonnell2021machine, otto2012multiphysical}. In the gradient search method, the mismatch between the achieved and desired machining qualities is minimized by changing the value of one processing parameter at a time. Gradient search is a deterministic optimization method and, thus, the search will terminate when the algorithm finds any minimum/maximum, whether local or global. As a result, gradient search also depends on the initial values and is highly unlikely to find the global optimum; therefore, gradient search does not eliminate the need to perform systematic experimental optimization. 

In the simulation method~\cite{otto2012multiphysical}, the complete laser machining process is modelled theoretically to determine the optimal combination of parameters. 
%\textcolor{red}
{This requires the ability to simulate realistic experiments of light-matter interactions which is a very complex task.}
%To fulfil this optimization, realistic experiments of light-matter interaction need to be simulated, and that leads to high computational complexity. 
It should also be noted that, oversimplification in such models through crude approximations can be detrimental to the simulation efficiency~\cite{mazhukin2017nanosecond}.

Machine learning (ML) is a new alternative to predict the behaviour of a complex system trained by a small experimental dataset~\cite{mcdonnell2021machine, feng2019using, zhang2018strategy, kechagias2022robust, chen2020critical, fatehi2021accelerated}. This strategy has been shown to be effective in modelling laser machining~\cite{mills2021lasers, heath2018single, heath2018machine, mills2018predictive, teixidor2015modeling, dhupal2007optimization}, and it is significantly faster in terms of computational speed. Different ML methods have been demonstrated to be able to model physical phenomena directly from experimental data without considering any underlying physical equations~\cite{mills2018predictive, xie2019deep, bakhtiyari2021review}. 
%\textcolor{red}
{
% In general, ML applications in laser machining studies are accomplished in three stages. Stage one is to collect experimental data containing the independent variables corresponding to laser machining outputs. Stage two is to train ML models with experimental data to predict the output parameters as a function of inputs. The third stage is to provide the trained model with a set of unseen data to predict their associated outputs~\cite{mills2018predictive}.
For instance, McDonnell et al.~\cite{mcdonnell2021machine} applied ML techniques to grey cast iron and studied the height of the laser produced crown and dimple depth as a function of the laser pulse energy, its repetition rate and the number of pulses. After comparing performance of different algorithms using the full range of each of the laser parameters, they achieved more accurate predictions using the neural network (NN) model. As another example, Teixidor et al.~\cite{teixidor2015modeling} used ML techniques on another metallic system (hardened steel) to create shallow microchannels.
%in the order of 20~$\mu$m. 
They studied the material removal rate as a function of scanning speed, laser pulse intensity and frequency, and found NN and decision trees algorithm to be more efficient 
%\textcolor{red}
{than their other tried models} for predicting channel geometry. Based on their experimental data, they declared the scanning speed and pulse intensity the most and the pulse frequency as the least relevant parameters for predicting depths.
As a final example, 
%On application of ML methods
%(NN technique) 
Dhupal et al.~\cite{dhupal2007optimization} applied ML to ceramics machining using aluminum titanate and an Nd:YAG laser. They investigated the effect of input parameters such as lamp current, laser pulse frequency and width, as well as the cutting speed and air pressure on creating microgrooves.
%Their selected ceramic system was aluminum titanate and the laser system was a Nd:YAG laser. 
Limited to a fixed focal position, they reported second order equations relating outputs to the input parameters to find the optimized inputs required for each target output. 
%The expected results through the optimization where within acceptable errors in respect to the NN predictions validated with experimental tests.
}

%\textcolor{red}
{In this work, ML techniques were employed to predict the dimensions of the engraved channels
% with triangular or trapezoidal cross-sections 
on alumina ceramics using ytterbium picosecond fiber laser. Different ML algorithms, including linear/polynomial regressions, XGBoost (XGB)~\cite{chen2016xgboost} which is a tree-based algorithm, and Multi-Layer Perceptron Neural Networks (NN)~\cite{Almeida1997-qo} methods, were compared based on their success in predicting the channel dimensions as a function of laser parameters. Based on the comparison, an NN-based model was developed to identify the laser processing parameters for fabrication of high-quality channel (cut) architectures of desired properties using industrial grade alumina ceramics. The performance criteria are the channels' depth, and top and bottom widths. Although the published literature contains studies on the use of ML techniques for material removal applications~\cite{mills2021lasers}, the number and scope of such studies on ceramics  are limited. Furthermore, unlike previous studies~\cite{dhupal2007optimization, mcdonnell2021machine}, the current work is not restricted to a single channel, but is extended to predicting dimensions of different channels with either triangular or trapezoidal cross-sections. Finally, a reverse prediction is presented as an example of a practical application, where target channels for a given application are manufactured by the assist of ML techniques in determining the required input parameters.}

%%%%%%%%%%%%%%%%%
% This paper is organized as follows. In Section~\ref{sec:methods}, the \textcolor{red}{model and } experimental procedure \sout{is}\textcolor{red}{are} briefly explained, and the interdependencies of the laser parameters are investigated. \sout{The performance of different ML algorithms in predicting channel depth is evaluated and XGB-determined hierarchy of laser parameters for predicting outputs (variable importance) is justified.} In Section~\ref{sec:ML} the focus is on the  ML models for predicting channel dimensions. \textcolor{red}{The performance of different ML algorithms in predicting channel depth is evaluated and XGB-determined hierarchy of laser parameters for predicting outputs (variable importance) is justified.} 
% Finally,  a model is chosen for a practical scenario (reverse prediction), where target channel geometries were selected and laser input parameters are predicted.\textcolor{red}{Input parameter combinations are produced via Generative Adversarial Network(GAN).}  The results of this practical evaluation are compared with experiments.
%%%%%%%%%%%%%%%%%

%\textcolor{red}
{This paper is organized as follows. In Section~\ref{sec:methods}, the model and experimental procedure are briefly explained, and the interdependencies of the laser parameters are investigated.  Section~\ref{sec:ML_methodology} focuses on ML methodology, which includes a brief explanation of the employed ML algorithms, preprocessing and model evaluation techniques. Also, performance of different ML algorithms is evaluated. The ML section continues with  XGB-determined hierarchy of laser parameters for predicting outputs (feature importance)
% in Section~\ref{sec:XGB}
and ends with exploration on the NN structure. 
% in Section~\ref{sec:NN}. 
Finally in Section~\ref{sec:application}, the performance of the NN model is evaluated by applying it to a practical scenario, where target channel geometries are selected and the laser input parameters are predicted (reverse prediction). Input parameter combinations are produced through Generative Adversarial Network (GAN)~\cite{goodfellow2014generative} and results of this practical evaluation are compared with experiments. The paper finishes with conclusions in Section~\ref{sec:conclusion}. Supplementary Information includes more detailed graphs of Figures~\ref{fig:amplitude} -~\ref{fig:frequency}, and the experimental data collected and used in this study.}
%%%%%%%%%%%%%%%%%%%%%%%%%%%%%%%%%%%%
\section{Model and experimental data} 
\label{sec:methods}

% \section{\sout{Methods and methodology }\textcolor{red}{Model and experimental data}\label{sec:methods}}

%\textcolor{red}
{
%The goal in this work is to employ ML techniques on experimentally driven laser machining data to find the parameter combinations required for engraving channels with target dimensions. 
In general, ML applications in laser machining studies are accomplished in three stages: 1.) Collect experimental data containing the independent variables corresponding to laser machining outputs, 2.) train ML models with experimental data to predict the output parameters as a function of inputs, and 3.) provide the trained model with a set of unseen data to predict their associated outputs~\cite{mills2018predictive}.}
%\textcolor{red}
{In this section, a brief explanation of the experimental setup and the procedure for data collection are explained. In addition, as preliminary analysis the effect of laser parameters on the channel dimensions are qualitatively investigated. ML models are discussed in Section~\ref{sec:ML_methodology}.
}

%%%%%%%%%%%%%%%%%%%%%%%%%%%%%%%%
\subsection{Experimental Procedure}\label{sec:Exp}

\begin{figure*}[h]
    \centering
    \includegraphics[width= 0.8\textwidth]{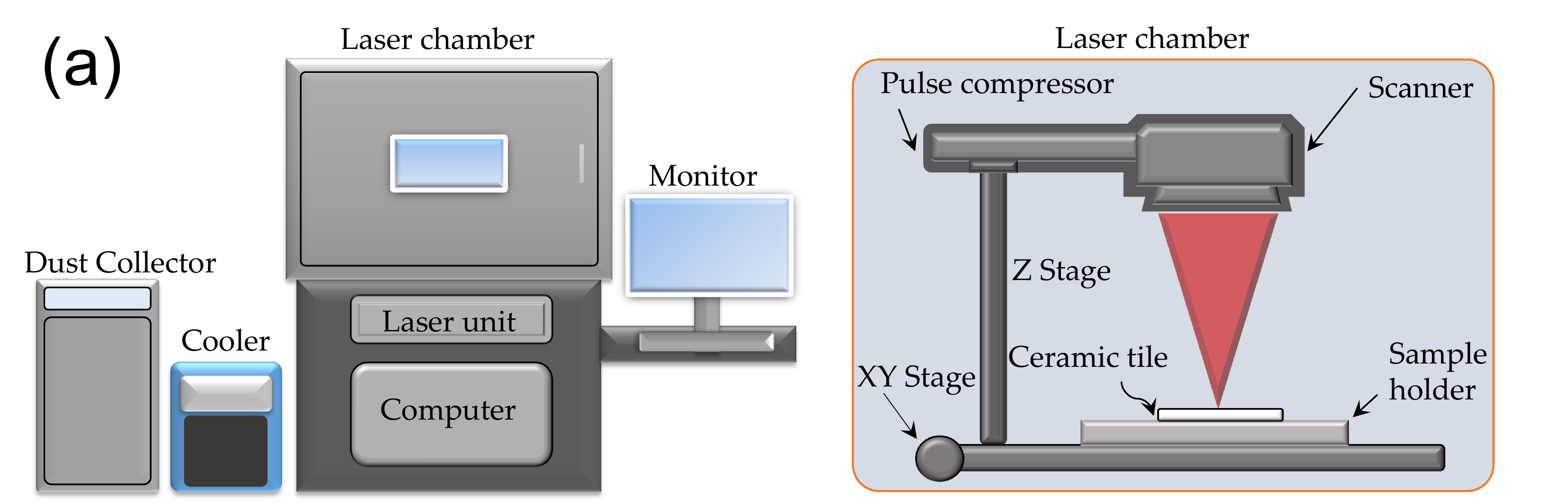}
    \includegraphics[width= 0.8\textwidth]{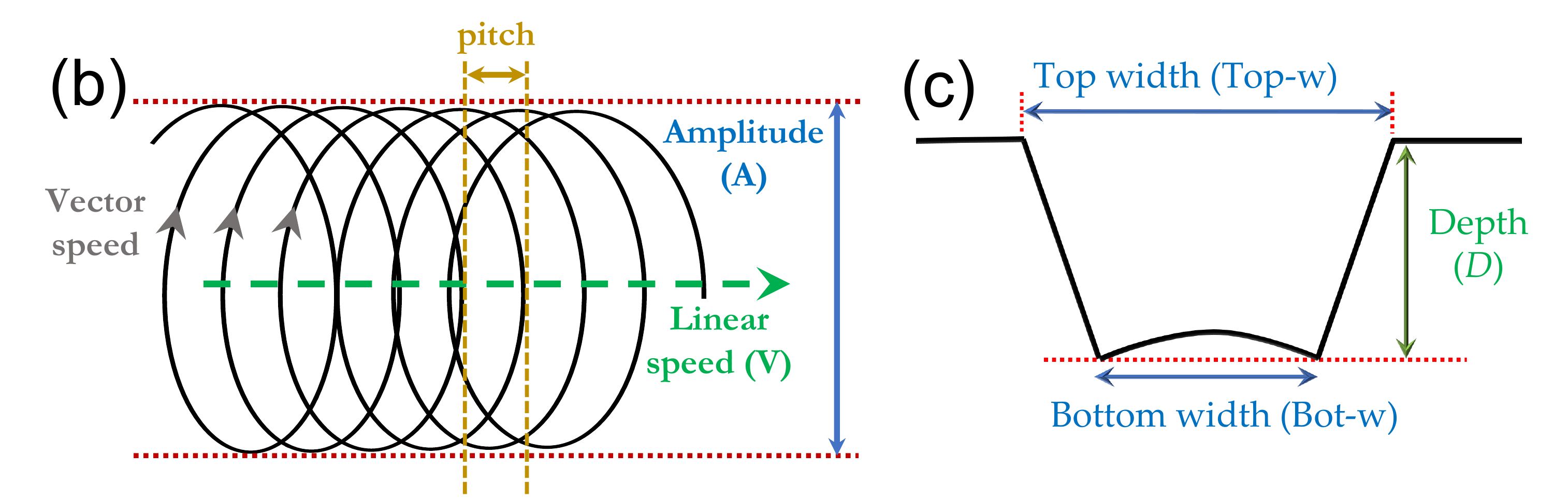}
    \caption{(a) Schematic of the laser system and the equipment. (b) Schematic of the circular wobble pattern illustrating the laser pulses and direction, wobble amplitude, wobble pitch, linear speed, and vector speed. (c) Schematic of a trapezoidal cut.
    The channel cross-section is triangular when the bottom width is negligible.
    }
    \label{fig:experiment}
\end{figure*}

The laser used in this study is an ytterbium picosecond fiber laser (YLPP-25-3-50-R, IPG Photonics, USA),
%\textcolor{red}
{schematically shown in Figure~\ref{fig:experiment}(a),} with a maximum average power of 50\,W and a laser beam waist diameter of 17~${\mu}$m. A Gaussian spatial profile beam with 3~ps long 25~$\mu$J pulses of 1030~nm wavelength is produced by the laser, which has a repetition rate of up to 1.83~MHz. 

Previous studies~\cite{beausoleil2020deep, esmail2021engineered} revealed that the optimal cut quality is achieved using a circular wobble pattern. This scheme produces an array of circular patterns parallel to the substrate surface, as shown in Figure~\ref{fig:experiment}(b) along with a visualization of the wobble pitch using a 1~mm amplitude. A schematic of the resulted channel geometry is shown in Figure~\ref{fig:experiment}(c), highlighting the important output parameters.  The laser process parameters that govern the geometry of this circular wobble pattern are listed in Table~\ref{tab:laser_params}. The ceramic used in this work was industrial nonporous 96\% alumina tiles (McMaster-Carr, CA) with a density and porosity of 3875~kg/m$^3$ and 0\%, respectively. The effects of the laser process parameters on the quality of the cuts were studied by generating a series of 10~mm long cuts. 
%\textcolor{red}
{First, the ceramic tile was placed in a fixture inside the chamber. Then, using the laser software, the focal position and other process parameters were set for laser machining.} Once processed, the ceramic tiles were cleaned using compressed argon gas, and wiped with ethanol to remove any residual debris from the sample surface and within the cut geometry. The cut depth as well as the top and bottom widths were then measured using a KEYENCE$^{\mathrm{TM}}$ 3D Laser Scanning Confocal Microscope. These measurements were done by choosing three regions along the cut line and averaging the surface profile by sampling 15 individual cross-sections that were spaced 2.84~${\mu}$m apart within each region.

\begin{table*}[h]
    \caption{Relevant variables for laser ablation.}
    \centering
    \footnotesize
    % \begin{tabular}[]{|p{3.0 cm}|p{2.0 cm}|p{4cm}|p{4 cm}|}
    \begin{tabular}{|c|c|c|c|}
    \hline 
        {\bf Parameter} & {\bf Symbol} & {\bf Description} & {\bf Variable Assignment \newline \& range of values} \\
    \hline \hline
        Linear speed (mm/s) & $V$ & Speed at which the tracking stage traverses & Dependent ($V = f/40$) \\
\hline        
        Amplitude (mm) & $A$ & The diameter of the circles in the wobble & Independent  [$0.100 - 1.200$]\\
        \hline
        Wobble frequency (Hz) & $f$ & Number of circular patterns per second & Independent  [$200 - 1600$]\\
        \hline
        Number of passes & $N$& Number of times the laser scans a line & Independent  [$10 - 160$] \\
        \hline
        Laser-substrate distance (mm)& $F_l$ & The distance between the laser and substrate surface & Independent  [$91.676 - 93.200$]\\
        \hline
    \end{tabular}
    \label{tab:laser_params}
\end{table*}

%%%%%%%%%%%%%%%%%%%%%%%%%%%%%%%%

\subsection{Parameter Dependence}\label{sec:param_dep}

Before applying any ML model, the dependence of channel dimensions, including the depth ($D$), and top (Top-w) and bottom widths (Bot-w), on the %\textcolor{red}
{tunable laser parameters of} wobble frequency ($f$), pulse amplitude ($A$), number of passes over the sample ($N$) and vertical distance of the laser scanner from the sample surface ($F_l$) were studied. The range of each parameter is provided in Table~\ref{tab:laser_params}. To have a more efficient laser machining process, the linear speed of scanning the sample surface ($V$) was set to be proportional to the beam frequency ($f$) making $V$ a dependent variable.

Our experimental dataset comprises of 124 different combinations of the above four independent input parameters 
%\textcolor{red}
{(as reported in Tables~SI-SV)}.
The influence of  varying the laser parameters 
on the channel dimensions are investigated in Figures~\ref{fig:amplitude}
-~\ref{fig:frequency}. Each graph shows groups of laser combinations in which three of the laser parameters were kept fixed and groups are distinguished with different colour symbols. 
%\textcolor{red}
{Clearly not all the colour coded groups cover the entire range of the changing variable.} These curves attest that there is no isolated and one-to-one correlation between the input and output parameters, that is, a comprehensive understanding of this non-linearly coupled problem is only achieved through a more in-depth study as explained hereafter. It is noteworthy that the influence of laser parameters on the outputs varies based on the values of the parameters, the surface material and the used laser machine, which can be different from other reported studies. 

%%%%%%%%%%%%%%%%%
\subsubsection{Amplitude}

\begin{figure*}
    \centering
    \includegraphics[width= 0.24\textwidth]{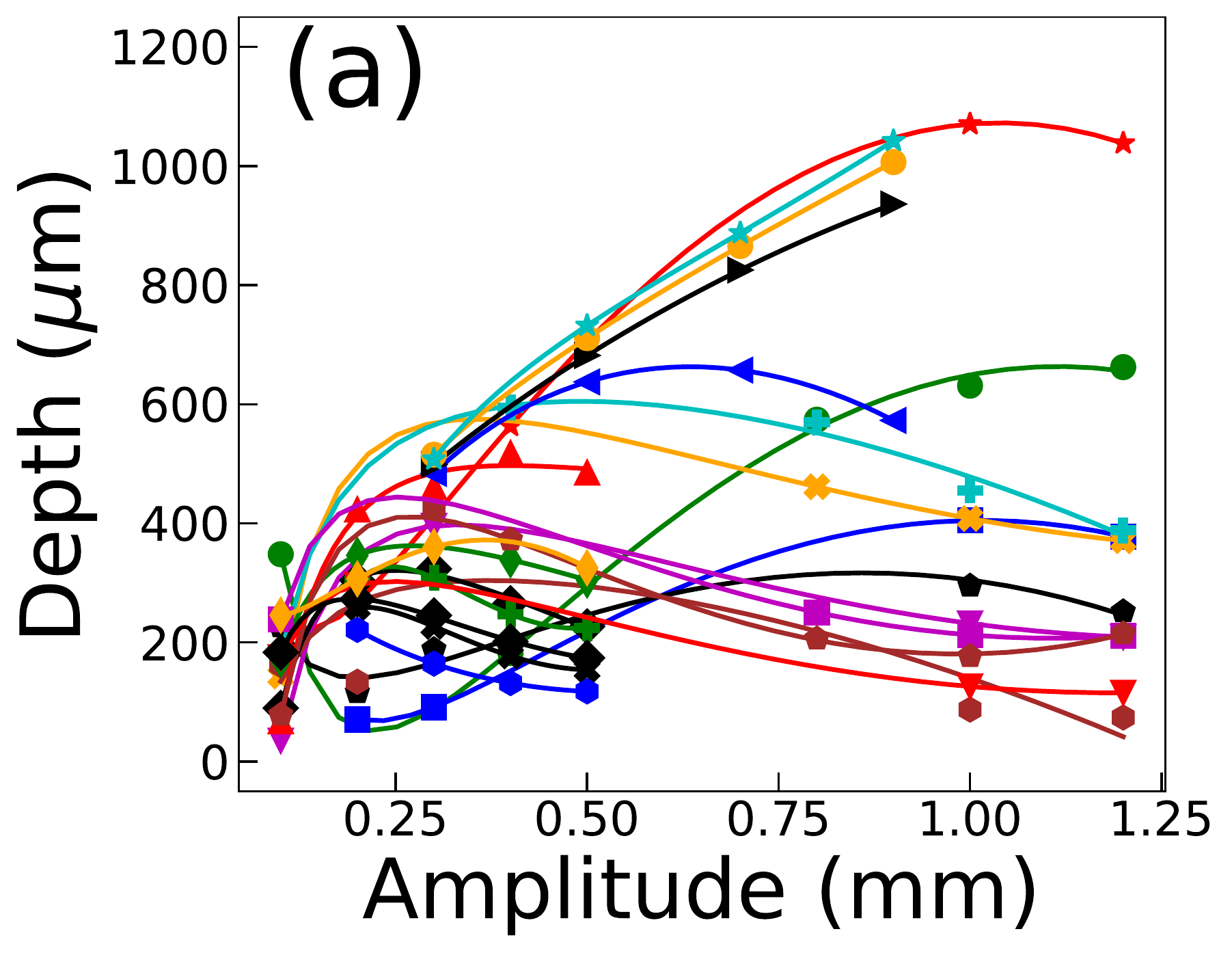}
    \includegraphics[width= 0.24\textwidth]{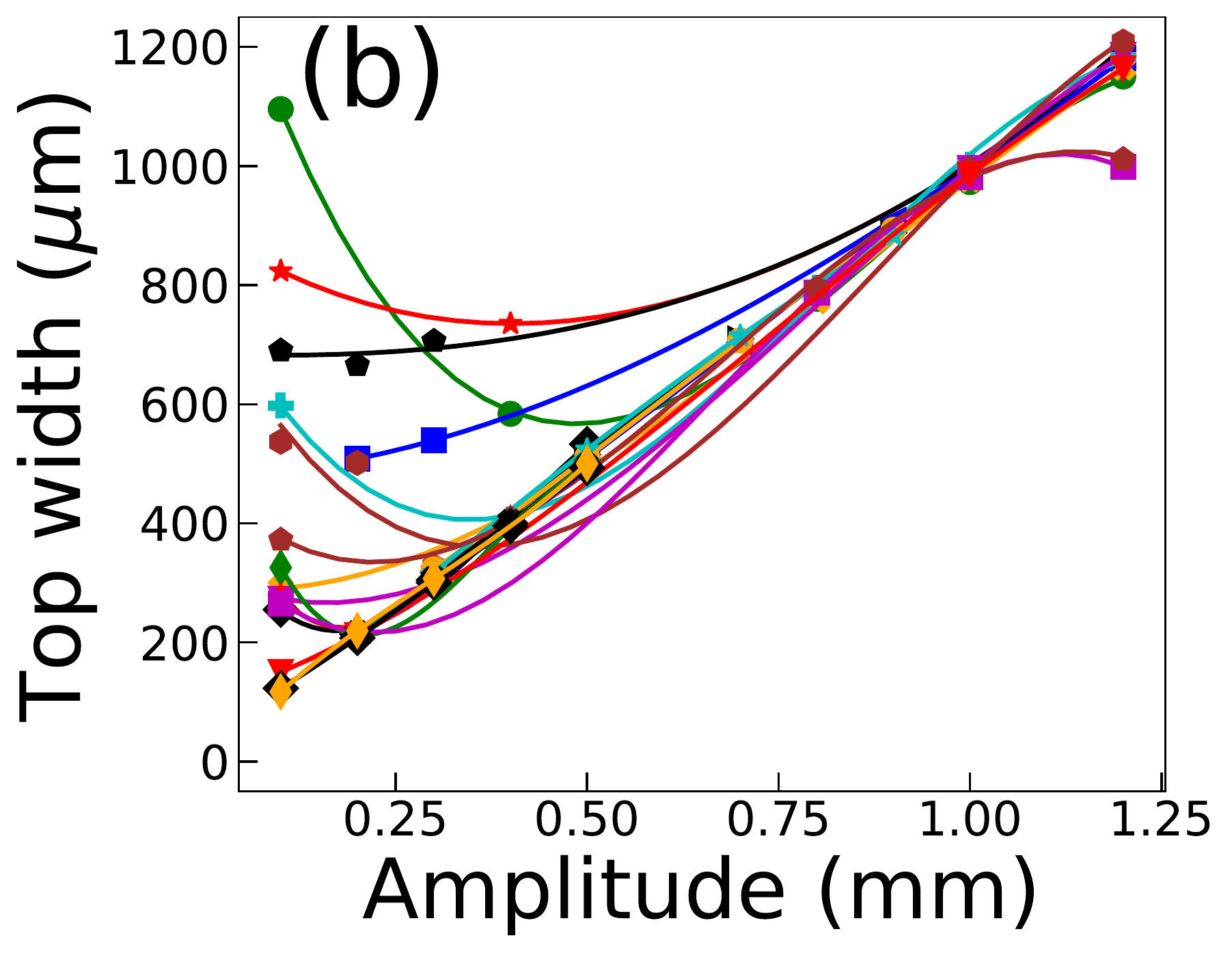}
    \includegraphics[width= 0.24\textwidth]{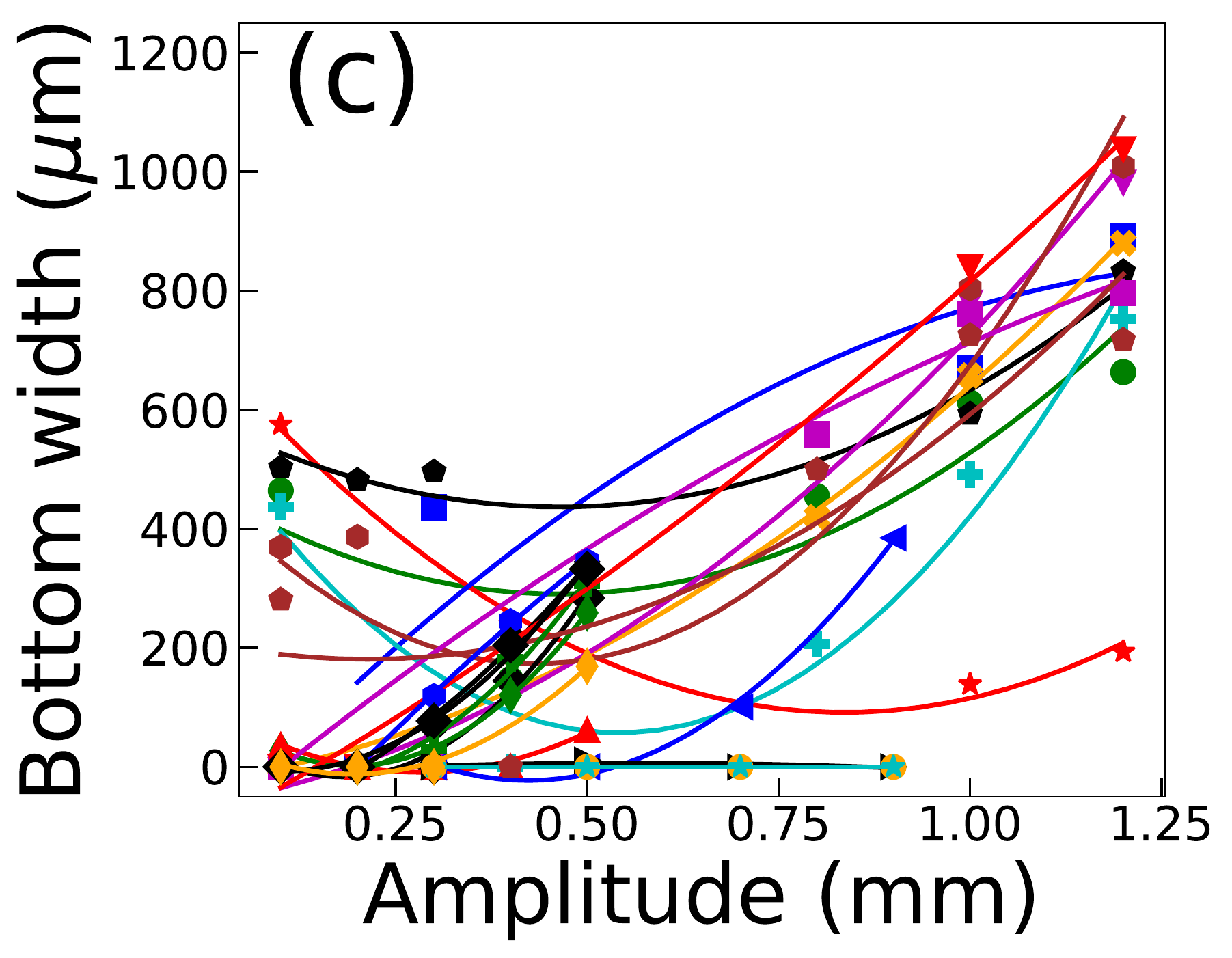}
    \includegraphics[width= 0.24\textwidth]{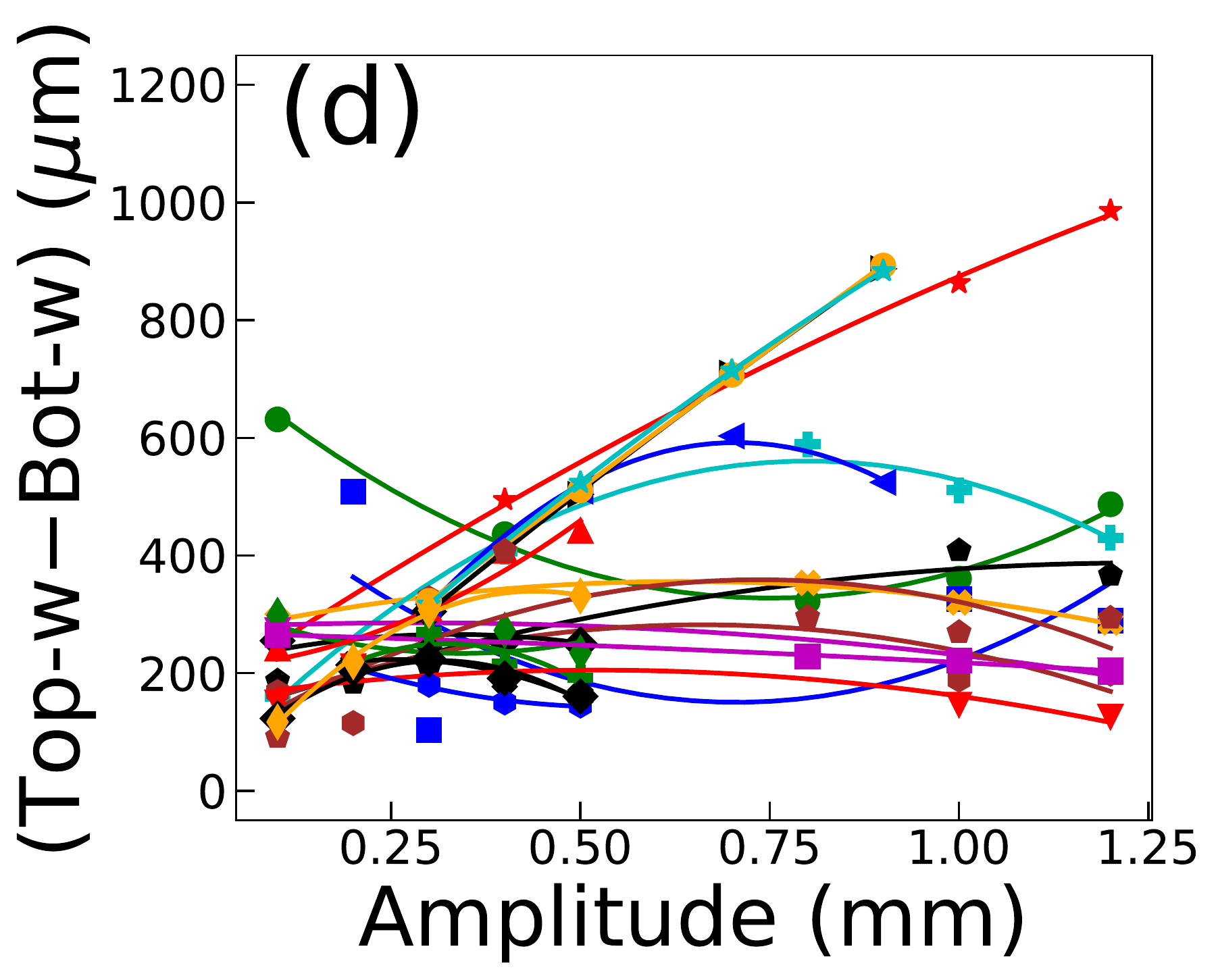}
    \caption{Dependence of the channel’s (a) depth, (b) top width, (c) bottom width and (d) the difference of the top and bottom widths on the laser beam amplitude. For depth, there is no consistent trend but for the top and bottom widths the behaviour follows a pattern. Each set of coloured symbols represents a group of experiments conducted with same values of $f$, $F_l$ and $N$ as reported in Figure~S1.}
    \label{fig:amplitude}
\end{figure*}

Figure~\ref{fig:amplitude} shows the effect of changing the pulse amplitude on the channels’ dimensions when the rest of the laser parameters are kept constant, as reported in Figure~S1. Figure~\ref{fig:amplitude}(a) demonstrates that the depth shows no consistent trend as a function of amplitude.
%, which means depending on other laser parameters, 
%increasing the amplitude can result in an increase 
%or decrease in the channel’s depth. 
For the top and bottom widths in Figures~\ref{fig:amplitude}(b) and (c), there is no linear relation between the amplitude and channel widths. In most groups, however, the width first decreases before starting to increase at an amplitude of around 0.5\,mm. This points to the key role of amplitude in determining the channel’s width. 
This is consistent with the results reported earlier~\cite{beausoleil2020deep, esmail2021engineered} 
%\textcolor{red}
{on change of trend for channel depth and width as a function of amplitude}. The differences of the top and bottom widths are more steady as a function of amplitude as shown in Figures~\ref{fig:amplitude}(d).

%%%%%%%%%%%%%%%%%%
\subsubsection{Number of passes}

\begin{figure*}
    \centering
    \includegraphics[width= 0.24\textwidth]{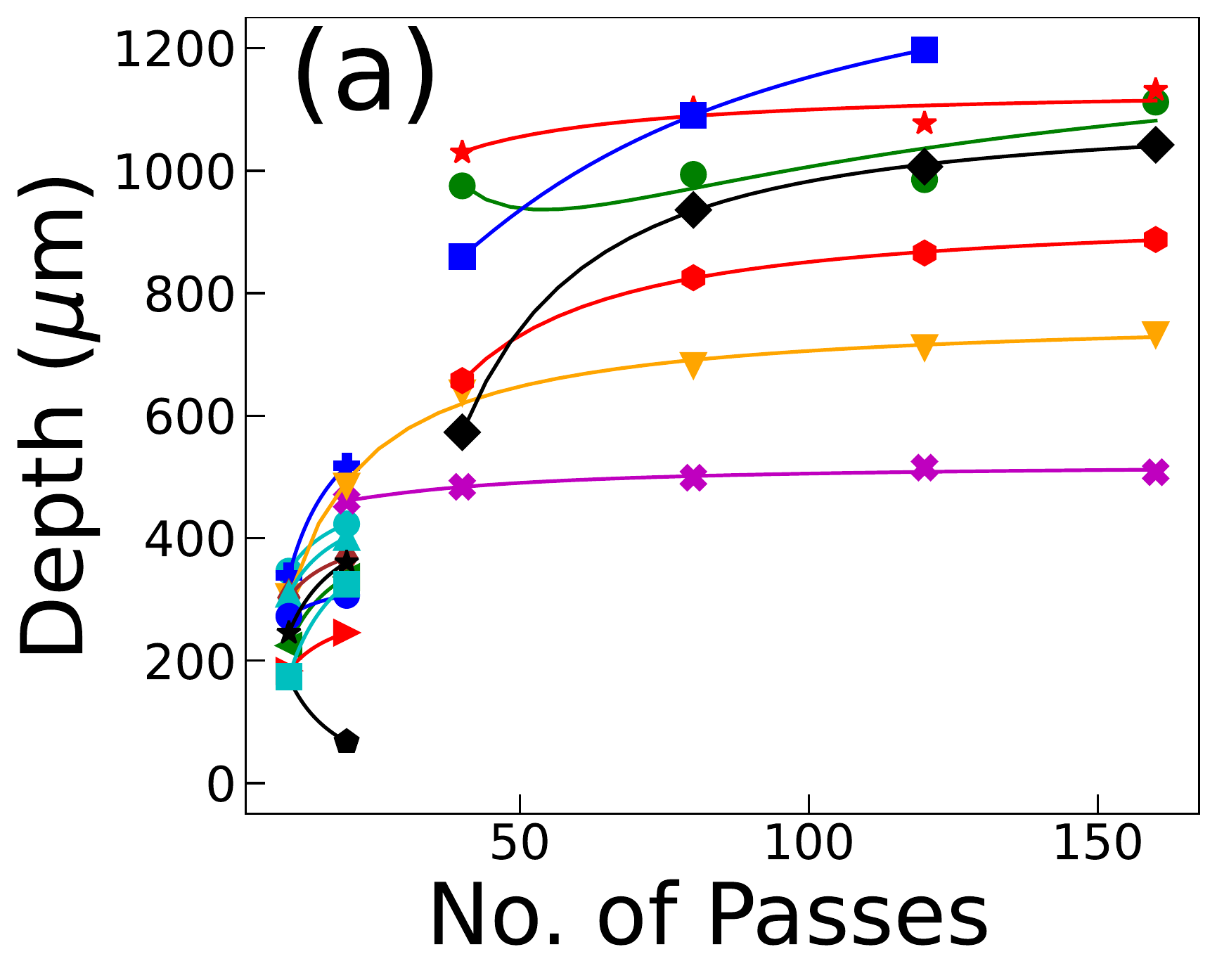}
    \includegraphics[width= 0.24\textwidth]{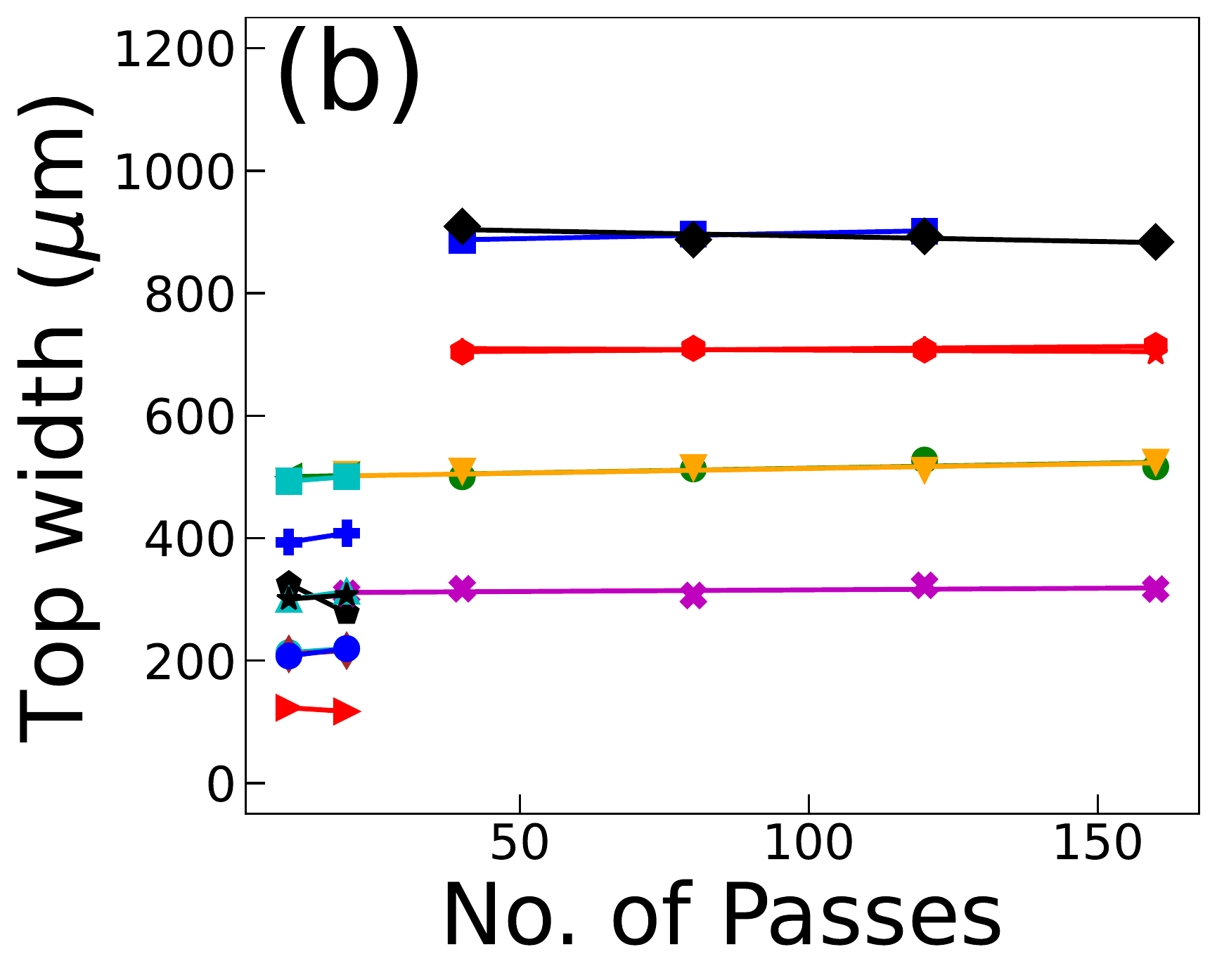}
    \includegraphics[width= 0.24\textwidth]{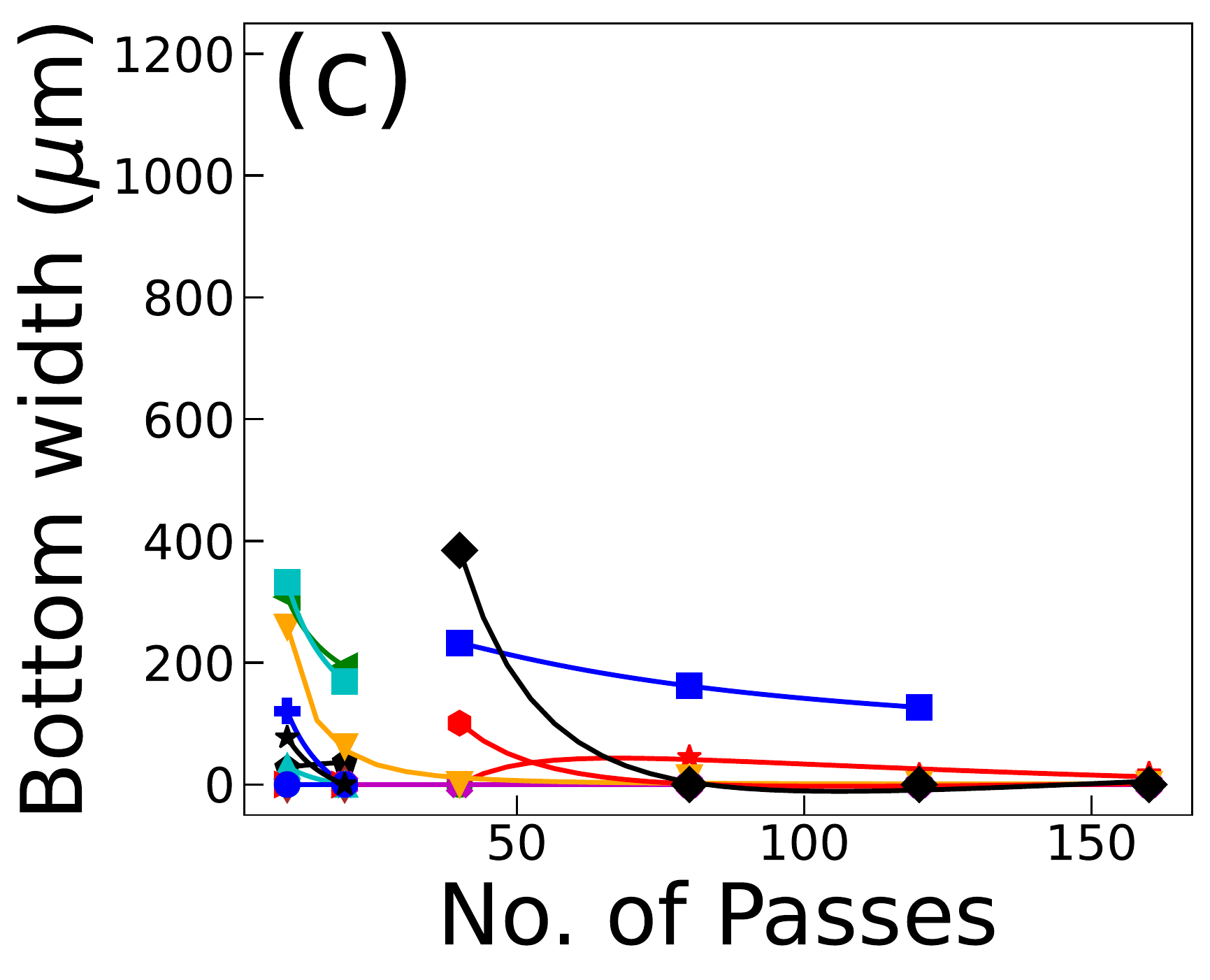}
    \includegraphics[width= 0.24\textwidth]{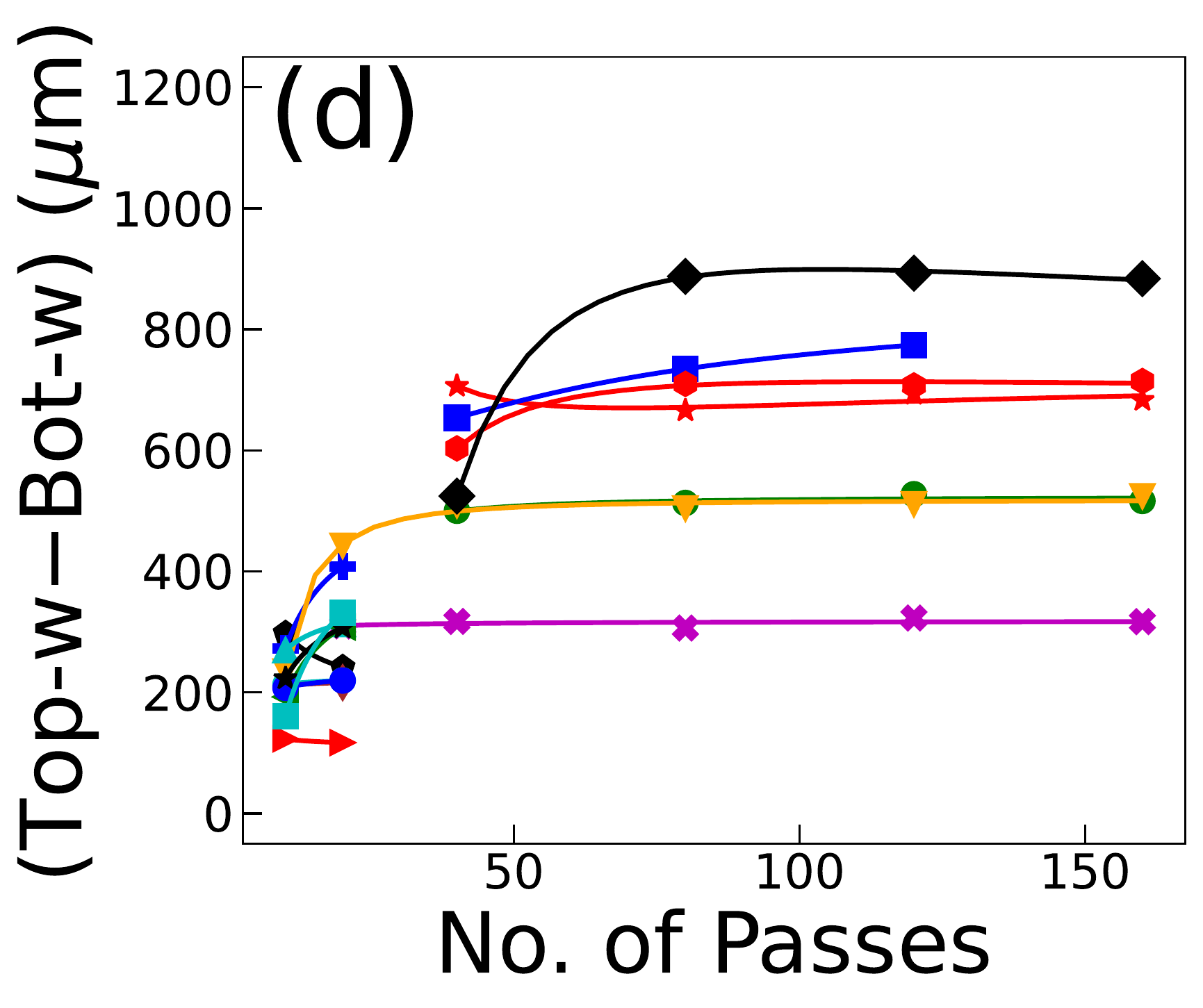}
    \caption{Dependence of the channel’s (a) depth, (b) top width (c) bottom width and (d) the difference of the top and bottom widths on the number of times the laser beam passes over the sample. Increasing the number of passes decreases the bottom width and can turn the trapezoidal channel into triangular. Each set of coloured symbols represent a group of experiments conducted with same values of $f$, $F_l$ and $A$ as reported in Figure~S2.}
        \label{fig:passes}
\end{figure*}

The results shown in Figure~\ref{fig:passes}(a) show an increase in the channel's depth when increasing the number of passes up to around 40, beyond which the trend reaches a plateau; 
%\textcolor{red}
{the top width (Figure~\ref{fig:passes}(b)), however, remains constant independent of the number of passes.}
%\textcolor{red}
{The behaviour of depth in Figure~\ref{fig:passes}(a)}
can be attributed to the fact that after some passes and cutting the target’s surface, less material remains in the laser’s focal point to cut, and the cut geometry turning into a triangular cross-section, along with possible material debris can contribute to a lower material removal rate. This conclusion is supported with the sudden decrease in the bottom widths when the number of passes exceeds 40 as shown in Figure~\ref{fig:passes}(c). This is in  agreement with the observations reported by Beausoleil et al.~\cite{beausoleil2020deep} 
%\textcolor{red}
{when they compare the channels' depth and cross-section shape as a function of number of passes}. Interestingly, the top width is almost independent of the number of passes and, as expected and shown in Figure~\ref{fig:passes}(d) owing to the abrupt decrease of the bottom width, the difference of the top and bottom width reaches a plateau beyond 50 passes.

%%%%%%%%%%%%%%%%
\subsubsection{Laser-substrate distance}
\begin{figure*}
    \centering
    \includegraphics[width= 0.24\textwidth]{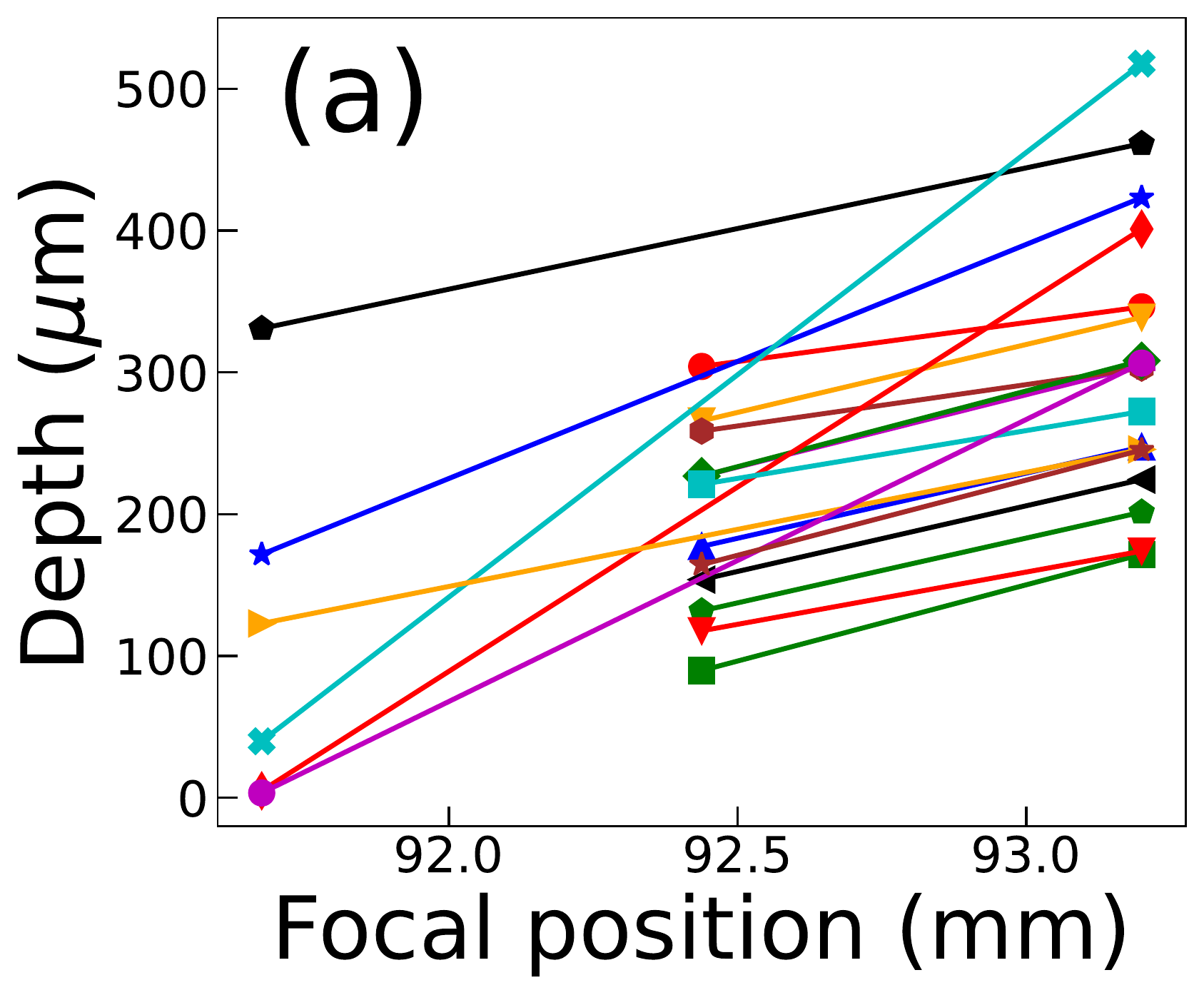}
    \includegraphics[width= 0.24\textwidth]{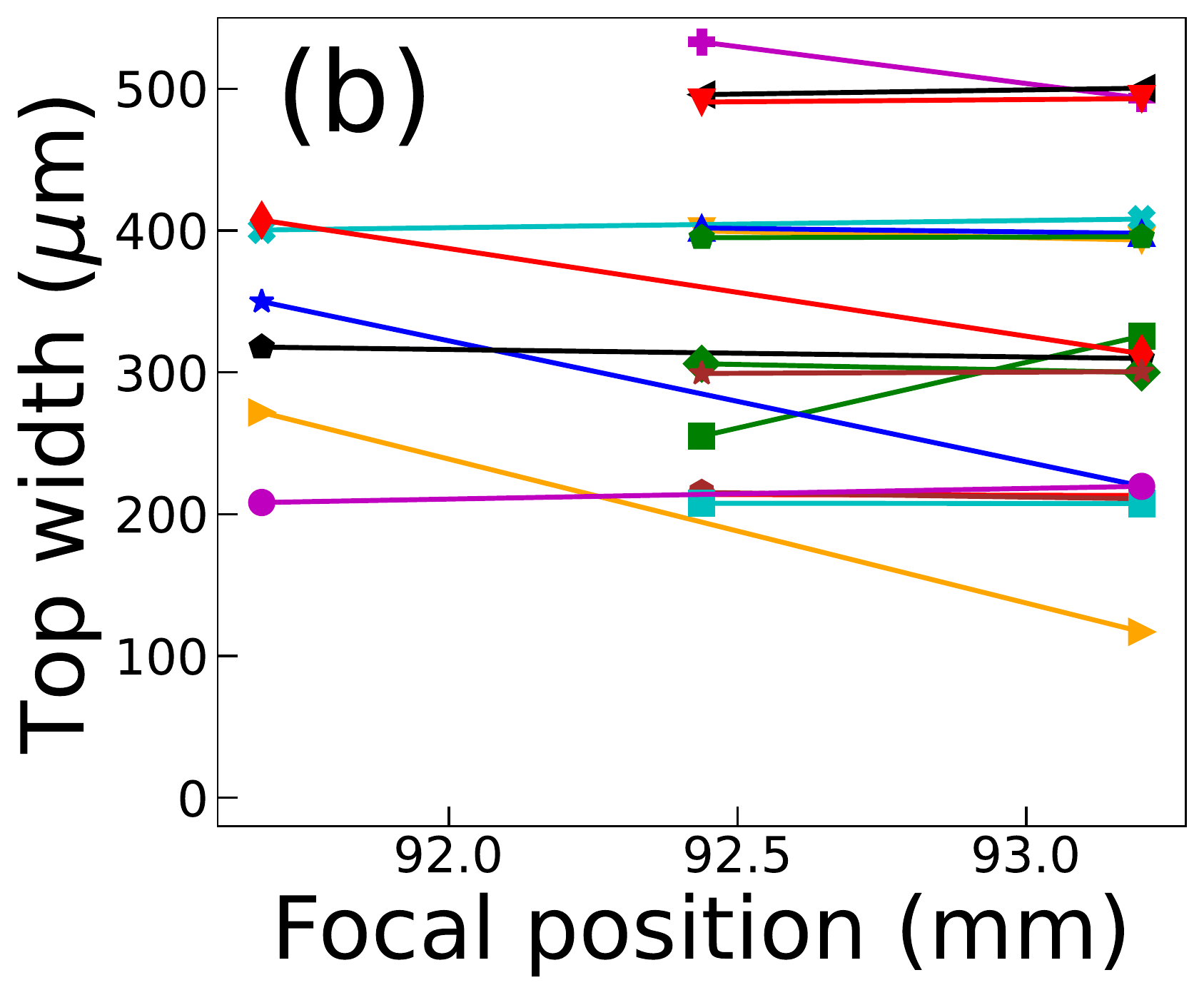}
    \includegraphics[width= 0.24\textwidth]{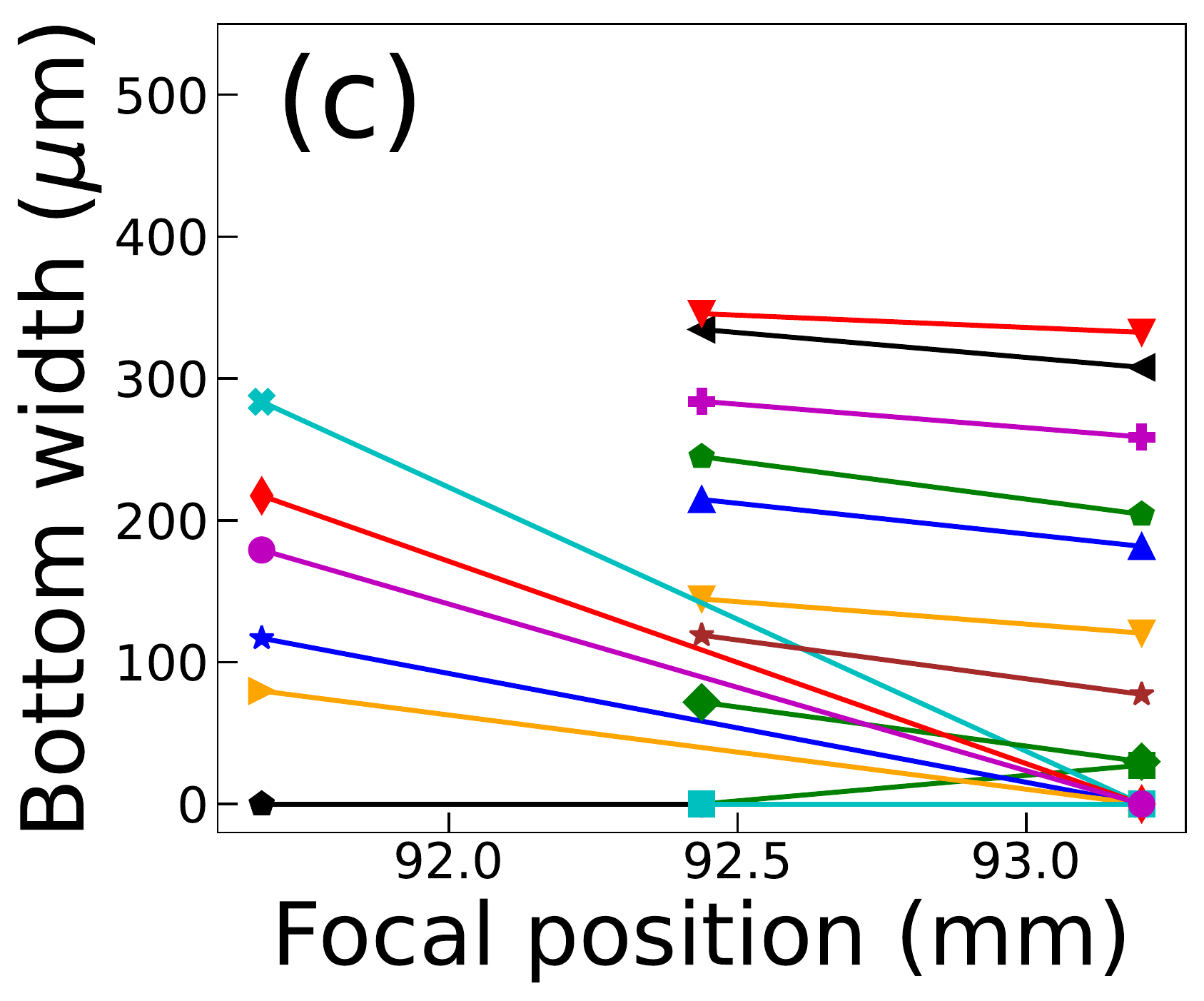}
    \includegraphics[width= 0.24\textwidth]{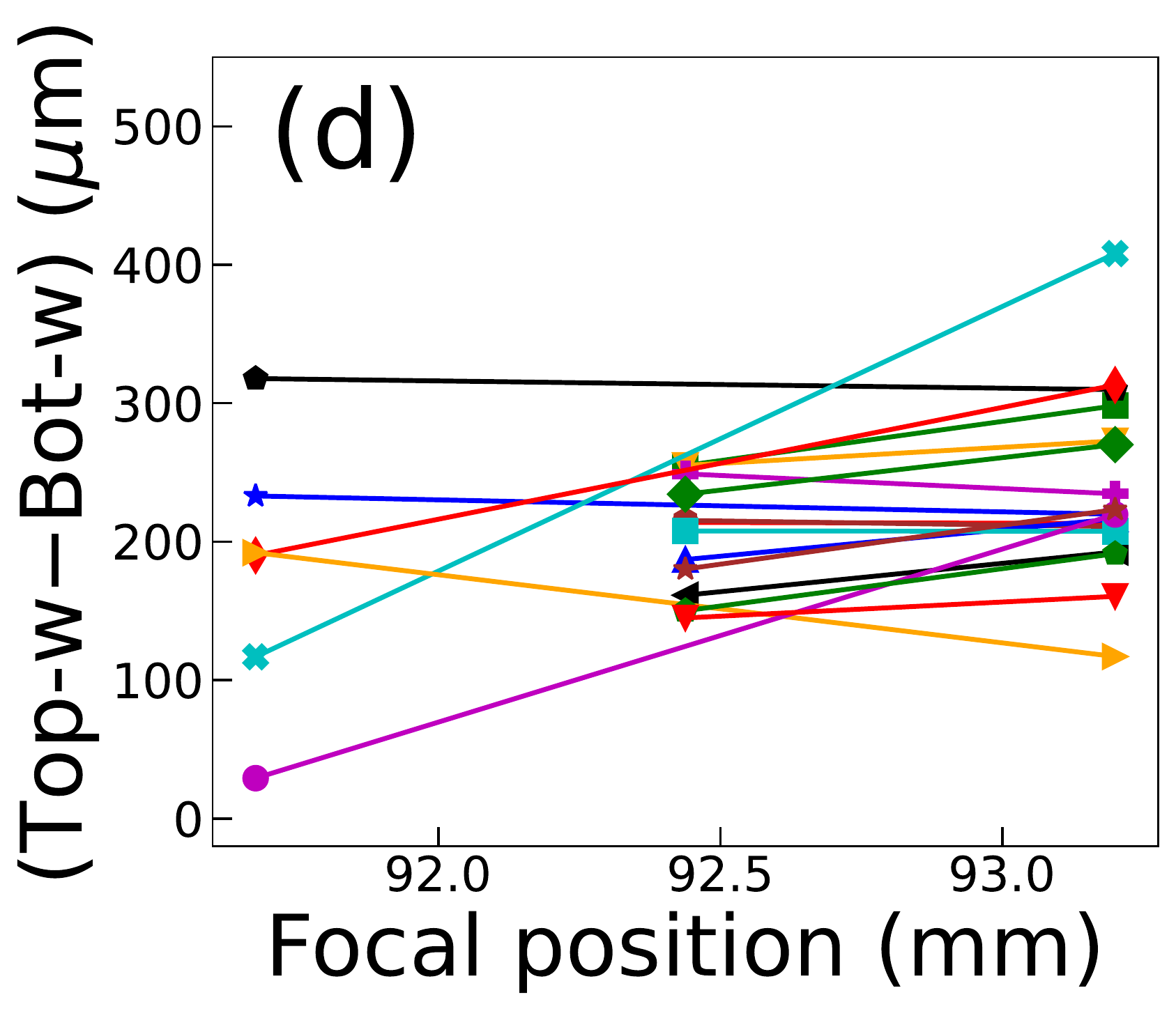}
    \caption{Dependence of the channel’s (a) depth, (b) top width, (c) bottom width and (d) the difference of the top and bottom widths on the vertical distance of the sample from the laser beam source ($F_l$). By fixing the sample at further distances from the beam source the channel’s depth increases and the bottom width decreases so that channel’s cross-section is closer to a triangle than a trapezoid. Each set of coloured symbols represent a group of experiments conducted with same values of $f$, $A$ and $N$ as reported in Figure~S3.}
    \label{fig:focal_position}
\end{figure*}

In this part, the influence of the laser-substrate distance 
($F_l$) 
on the channel dimensions is investigated. Interestingly, deeper channels are formed when the laser is further away from the sample’s surface, see Figure~\ref{fig:focal_position}(a). This can be understood by considering the Gaussian profile of the applied laser beam. As discussed by Esmail et al.~\cite{esmail2021engineered}, the depth and widths of the channel depend on the vertical distance of the laser source from the sample’s surface; 
they investigated the channel’s dimensions by fixing the laser source at six different distance levels from the sample’s surface and illustrated how the channel dimensions change when this distance deviates from the optimum value. If the optimum distance is defined for the position in which the material removal is maximised, starting a cut on a pristine surface with the optimum distance will result in optimal material removal rate at first. However, the active front of material removal changes with the removal of further material (as the channel progressively deepens), and that results in deviations from the optimum material removal rate. 
In the current study, the actual focal point of the laser system is set based on the substrate material, and therefore, an estimation of the optimum laser-substrate distance was known. The values presented here vary slightly around the optimum distance. It should also be noted that the changes applied to the laser-substrate distance are comparable in magnitude to the range of the desired channel depths, so the laser focal point is typically somewhere between the top of the surface and the bottom of the channel. 
Deeper channels were achieved at higher values of the laser-substrate distance, where the laser’s focal point was closer to the substrate's surface. Unlike the depth, the overall impact of the laser-substrate distance on the channel’s top width is not strong (see Figure~\ref{fig:focal_position}(b)), whereas the bottom width decreases with increasing this distance (see Figure~\ref{fig:focal_position}(c)). Putting these facts together, one can expect deeper triangular channels by fixing the focal point between the top and bottom surface of the substrate, as opposed to the case for shallow trapezoids. The difference of the top and bottom widths of the engraved channels mostly changes within $100-300~\mu$m range as shown in Figure~\ref{fig:focal_position}(d). The laser parameters employed for each colour group are reported in Figure~S3.

%%%%%%%%%%%%%%
\subsubsection{Frequency}

\begin{figure*}
    \centering
    \includegraphics[width= 0.24\textwidth]{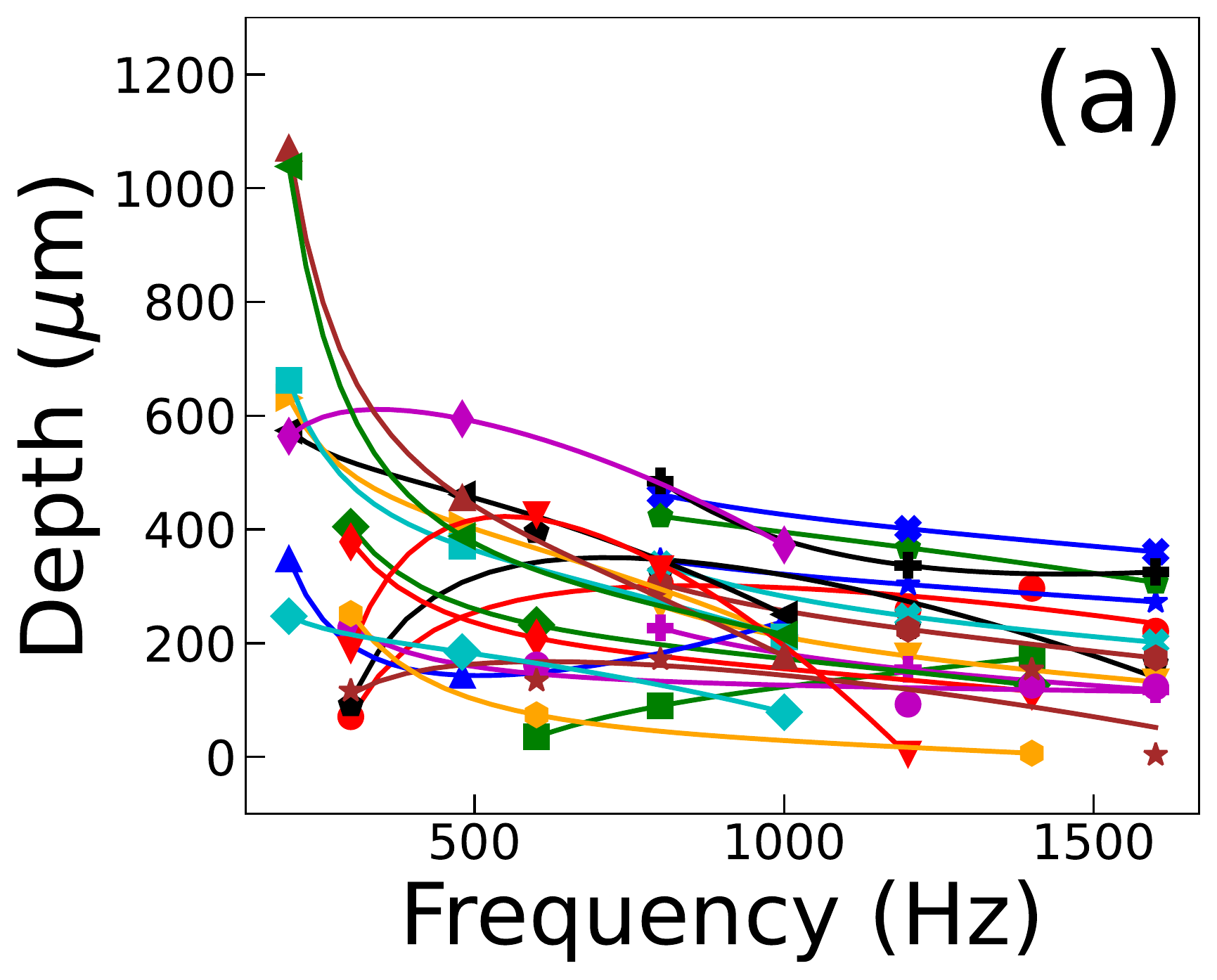}
    \includegraphics[width= 0.24\textwidth]{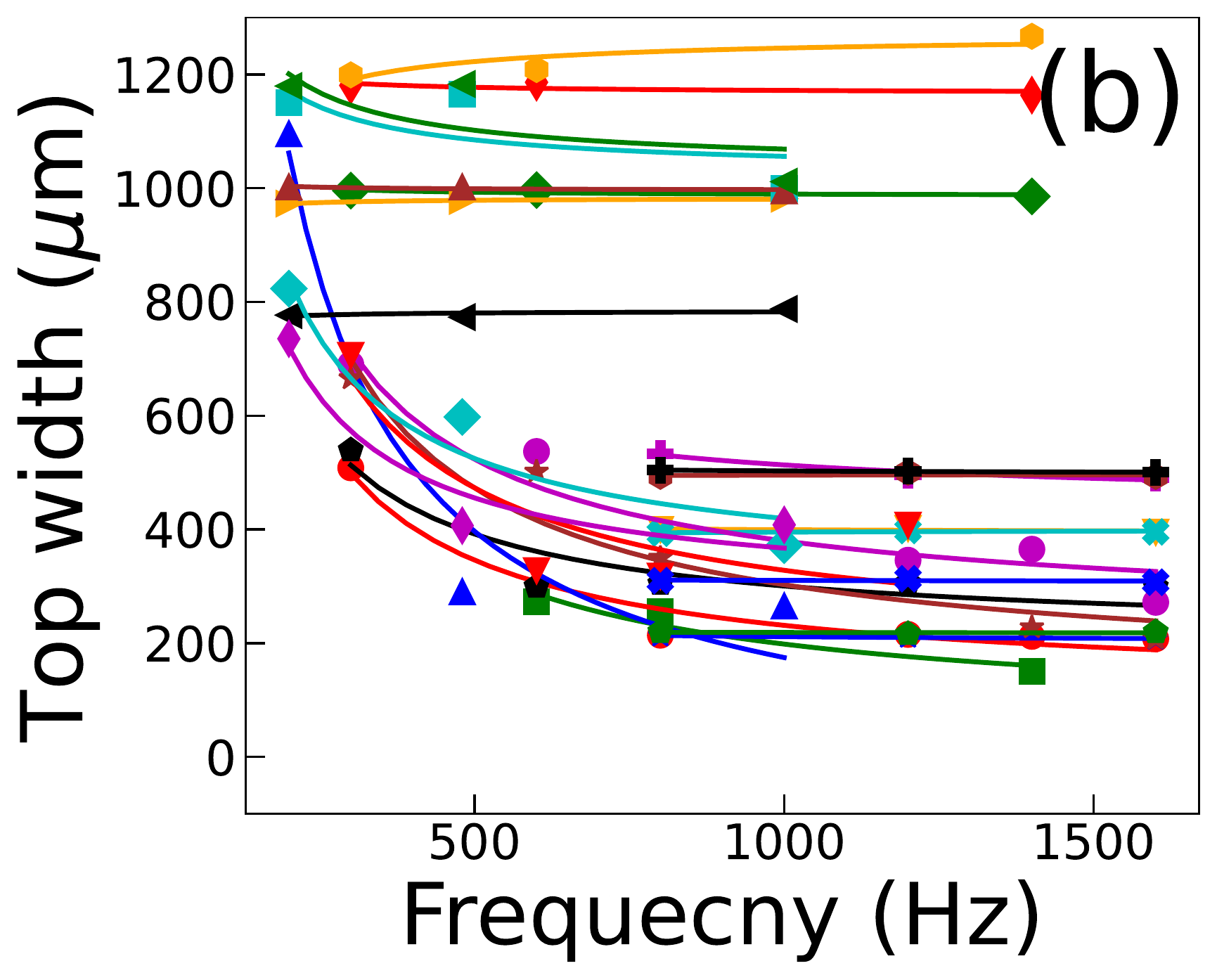}
    \includegraphics[width= 0.24\textwidth]{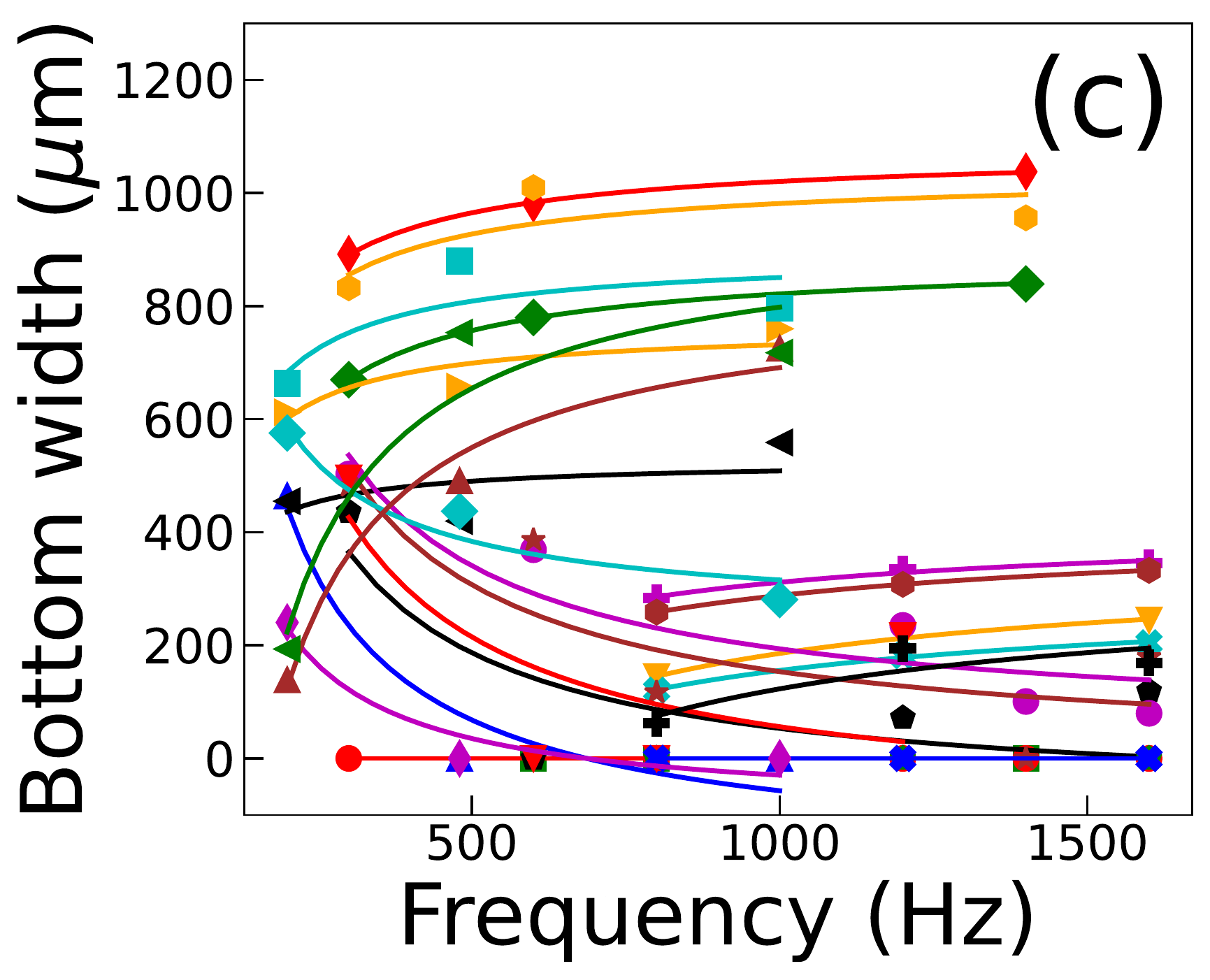}
    \includegraphics[width= 0.24\textwidth]{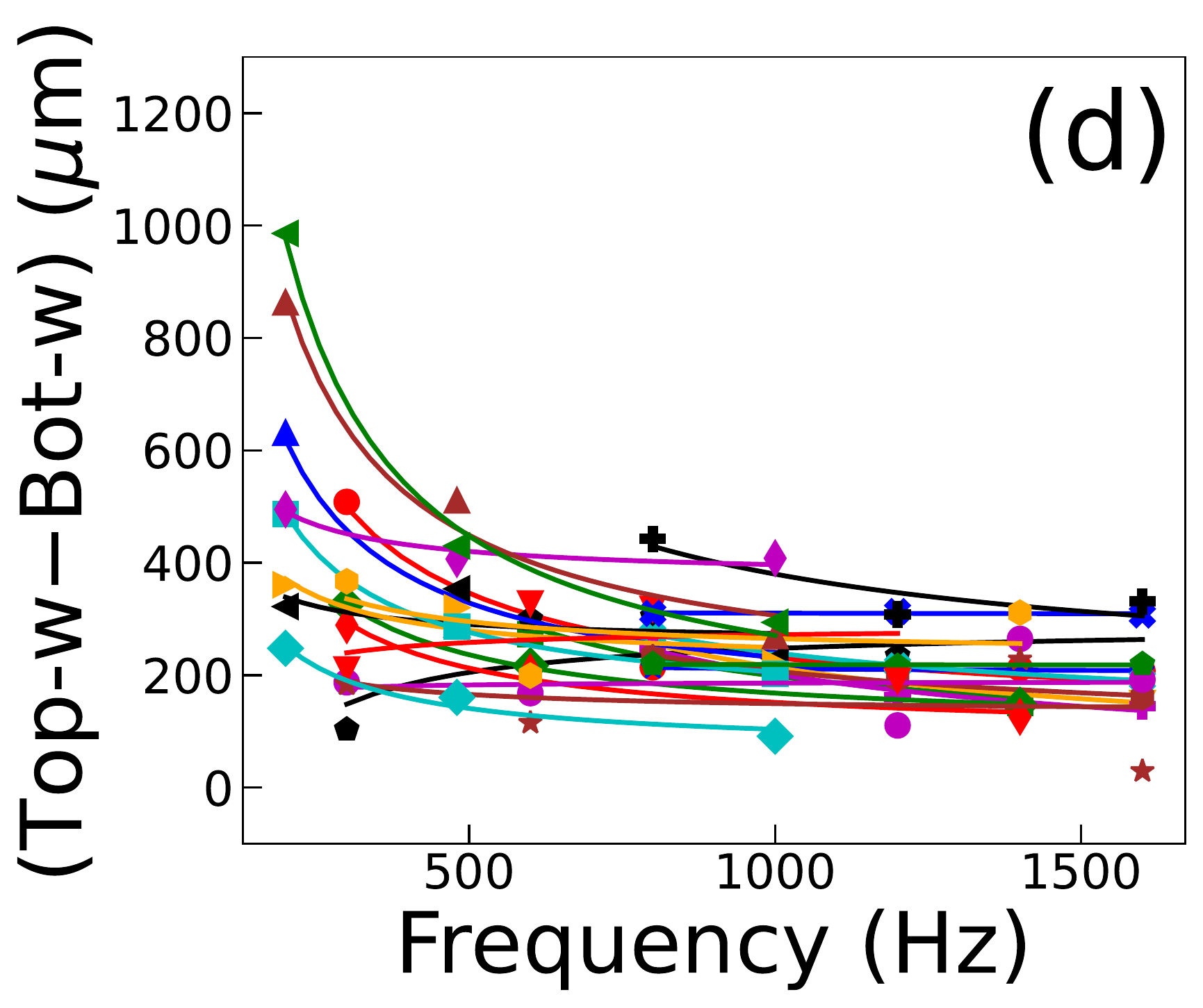}
    \caption{Dependence of channel’s (a) depth, (b) top width, (c) bottom width and (d) the difference of the top and bottom widths on the frequency of the laser beam. Each set of coloured symbols represent a group of experiments conducted with fixed values of $A$, $F_l$ and $N$, which are reported in Figure~S4.}
    \label{fig:frequency}
\end{figure*}

The change of channel dimensions as a function of frequency (or equivalently speed) is illustrated in Figure~\ref{fig:frequency}. The overall trend, 
as shown in Figures~\ref{fig:frequency}(a) and (b), is decreasing for frequencies lower than 500~Hz.
This is consistent with the results in Refs.~\cite{esmail2021engineered, teixidor2015modeling} 
%\textcolor{red}
{which report a decreasing trend of channel depth as a function of frequency or linear speed}. In Figure~\ref{fig:frequency}(c), the bottom width shows no consistent trend as a function of frequency. This means that depending on the values of other parameters, increasing the frequency can result in an increase or decrease in the channel’s bottom width. Figure~\ref{fig:frequency}(d) shows the differences between the top and bottom widths as a function of frequency, and one can conclude that for channels with trapezoidal cross-sections (where the bottom width is not zero), increasing the frequency results in smaller differences.
For cases with similar depths that means steeper side-walls. 

%%%%%%%%%%%%%%%%%%%%%%%%%%%%%%%
\section{ML methodology}\label{sec:ML_methodology}

%\textcolor{red}
{
%As an introduction to the second step, which is the  
% training of an ML model with the experimental data, 
This section provides a brief explanation of the employed ML models, data preprocessing and model evaluation techniques.
%Second step proceeds with comparing model performances for predicting channel dimensions. The feature importance for each output prediction and NN structure exploration are the complementary discussions.
} 

\subsection{Algorithms}\label{sec:algorithm}
The ML algorithms used in this work are linear and polynomial regression models~\cite{schneider2010linear}, a tree-based model called XGBoost (XGB)~\cite{chen2016xgboost}, and Multi-Layer Perceptron NN~\cite{bishop1995neural}. They are discussed below.

\subsubsection{Regression}

%\textcolor{red}
{Regression analysis is commonly used as the primary method for assessing relationships between data inputs and outputs through both linear or non-linear analysis (e.g. using higher order polynomials)~\cite{mills2021lasers}.}
In regression (Reg.) models~\cite{herbrich1999support, schneider2010linear}, a target parameter $Y$ is predicted by a linear combination of weighted input parameters (so-called predictors), $X_1$, $X_2$, ..., $X_n$ and their interactions
in which the weights ($w_i$) and intercept ($b$) are determined such that
the mean squared error (MSE) of the training data is minimized. 
The relationships of three predictors 
can be described as

\begin{widetext}
\small
\begin{equation}
\label{eq:poly_regression}
%\scalebox{0.85}%{%
\begin{split}
\!\! \mathrm {Linear Reg.}(Y)& \!= \!w_1 X_1 \! + \! w_2 X_2 \! + \! w_3 X_3 \! +\! w_4 X_1 X_2 \! +\!  w_5 X_1 X_3 \! + \! w_6 X_2 X_3 \! + \! b\\
\!\!\ \mathcal{O}(2)\mathrm {Poly. Reg.}(Y)& \! =\! \mathrm{LinearReg.}(Y) + w_7X_1^2+w_8X_2^2+w_9X_3^2\\
\!\!\ \mathcal{O}(3)\mathrm {Poly.Reg.}(Y)& \!= \! \mathcal{O}(2)\mathrm {Poly.Reg.}(Y)+w_{10}X_1^3+w_{11}X_2^3+w_{12}X_3^3, \\[6pt]
\end{split}
%}
\end{equation}

\end{widetext}

\normalsize
where, in this case, $Y$ can be the channel’s depth, top or bottom width and the $X_j$s are the laser input parameters
(wobble frequency, beam amplitude, number of passes of the laser over the surface, and the laser source’s vertical distance from the top surface). Higher order polynomial regressions follow the same trend as
the set of 
Equations~\ref{eq:poly_regression} by adding higher powers of parameters to the lower order terms. 

\subsubsection{XGBoost}\label{sec:XGB}

XGBoost (XGB)~\cite{chen2016xgboost}, Extreme Gradient Boosting, is a decision tree ensemble~\cite{Opitz1999-nn}. 
%% \textcolor{red}{Decision trees are a type of supervised learning algorithms based on partitioning data into subsets. Each Split is based on a particular variable, called a node, and in a specific location.}
Decision trees are a type of supervised learning algorithms in which the data are constantly partitioned into subsets. 
%\sout{split according to the values of the input parameters.} 
%\sout{Trees are exhibited with nodes and leaves. Nodes are the input parameters upon which the decisions are based, and leaves are the final outcomes of the splits.}
%\textcolor{red}
{Each data split is based on a particular variable, called a node, and in a specific location, which is the value of that variable.}
% {Splits are}
%\sout{Each split, determined based on what node and what value of the node, is} attributed with a score representing the amount of information gained (IG) or improved performance.  Different algorithms can be used for finding a hierarchy of the splits with the highest possible information gain and lowest error in \sout{the leaves} 
% \textcolor{red}
% {predicted outputs}.
XGB is a tree ensemble model that finds the optimized tree by summing the predictions of multiple trees. Successive decision trees are applied to the same dataset with the objective of optimizing the previous trials and minimizing the errors~\cite{chen2016xgboost}.   

Feature importance through XGB is determined based on the gained information (IG score) attributed to each split node, which provides a sorted list from  the highest to the least relevant parameters for predicting an output. Feature importance is  discussed in Section~\ref{sec:feature_importance}.   

\subsubsection{Neural networks \label{sec:NN}}

\begin{figure*}
    \centering
    \includegraphics[width= 0.75\textwidth]{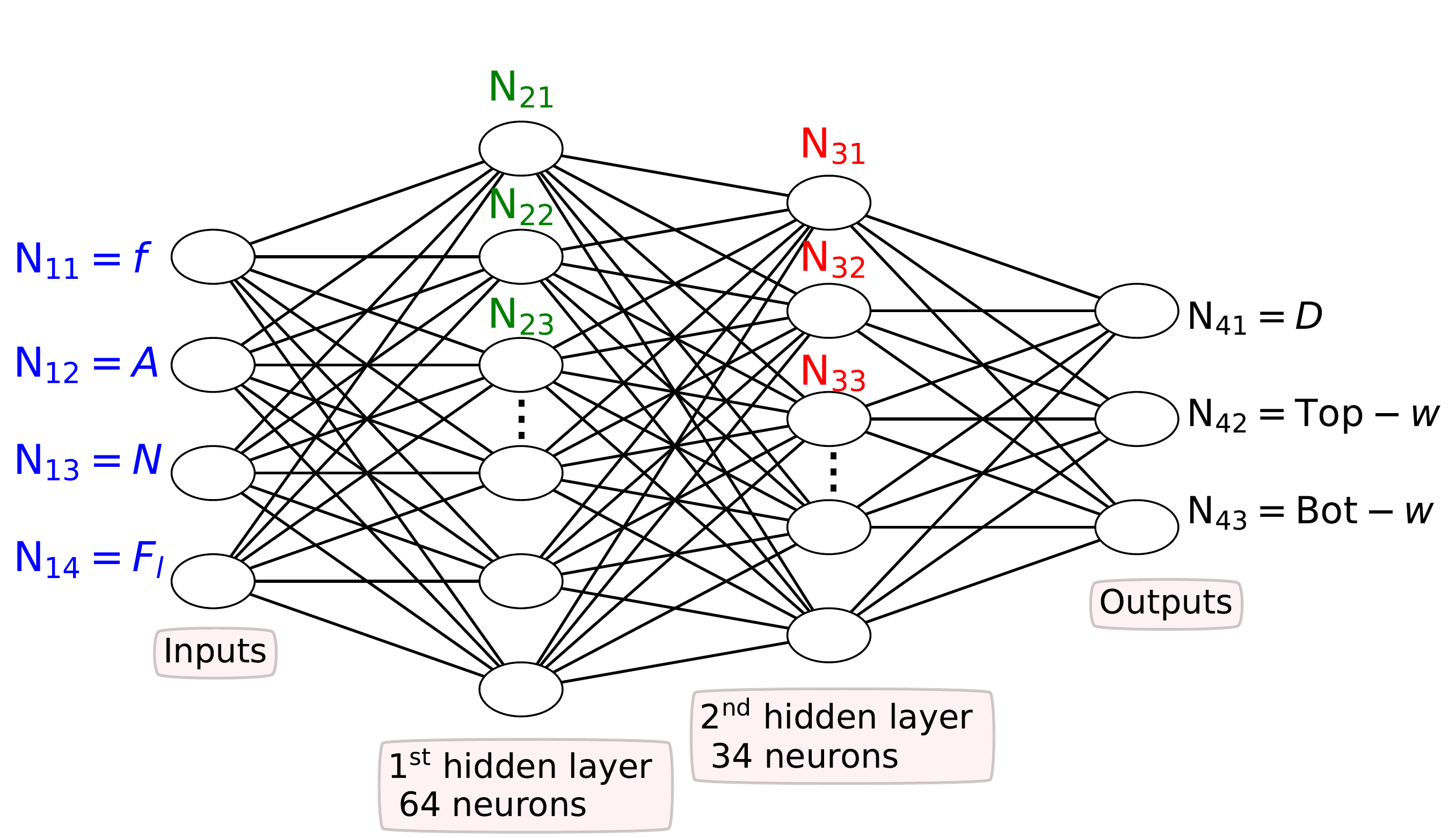}
    \caption{Schematic structure of a neural network connecting the input layer neurons of laser parameters such as the beam frequency ($f$), amplitude ($A$), number of passes ($N$) and laser-substrate distance
    ($F_l$) to the outputs of depth ($D$), top (Top-$w$) and bottom widths (Bot-$w$) of the channel through two hidden layers of 64 and 32 neurons.}
    \label{fig:nn}
\end{figure*}

Artificial neural networks (NNs)~\cite{Schmidhuber2015-uv} are mathematical models that allow complex relationship between input and output parameters. NNs are characterized by neurons connecting the independent parameters in the input layer to dependent parameters in the output layer through neurons in hidden layers, see Figure~\ref{fig:nn}. Each neuron ($N_{lk}$) in a hidden layer ($l$) is a nonlinear function ($f(z_l)$, called activation function) of a linear combination ($z_l$) of the neurons in the preceding layer ($l-1$),
\begin{equation}\label{eq:nn}
\begin{split}
    N_{lk}& = f(z_{l})\\
    z_l & = \displaystyle \sum_m w_m N_{l-1m} + b,
\end{split}
\end{equation}
where $w_m$'s are the weights and $b$ is a constant, called bias, in the linear term. The sum is over the number of neurons in the preceding layer,
%shown with 
$N_{l-1m}$. The activation function $f(z)$ can be in the form of an identity function, logistic  function (sigmoid), hyperbolic tangent ($\tanh$), rectified linear unit (ReLU) or leaky ReLU. Details about these activation functions can be found in Ref.~\cite{Feng2019-yj}. 

In the current study, an NN algorithm connects four independent laser parameters to the channel's %\sout{dimensions}
%\textcolor{red}
{depth, top width and bottom width} via two hidden layers with 64 and 32 neurons (the notation for this is 4/64/32/3). This is a feed-forward structure in which each layer’s neurons are fully connected to the neurons in the adjacent layers, as shown in  Figure~\ref{fig:nn}. Weights are initialized randomly and are gradually modified in a back-propagation scheme to reduce the loss function 
%\textcolor{red}
{(%MSE here as is 
defined in the next section)}~\cite{marquardt2021machine}.
Since the NN predictions vary as a function of the weights' initial values, the NN outputs hereafter are shown within an error interval determined with the standard deviation of the predictions through 100 different initialization.  

%%%%%%%%%%%%%%%%%%%%%%%%%%%%%%%%

\subsection{Preprocessing techniques}\label{sec:preprocessing}

As shown and discussed in Section~\ref{sec:param_dep}, the correlations between the laser parameters and the channel dimensions are complicated. Therefore, ML techniques are employed to predict the set of laser parameters for  creating a channel with desired dimensions. 
The MSE,
%\textcolor{red}
{
the loss function,} and R-squared ($R^2$ score),
%\textcolor{red}
{representative of the model performance,} are defined below in Equations~\ref{eq:MSE} and \ref{eq:R2}, and are reported here to establish a comparison between different ML algorithms and to evaluate their accuracy in predicting channel dimensions ($y$ represents either $D$, Top-w or Bot-w) as functions of independent variables (e.g. $f$, $A$, $N$ and $F_{l}$). 

The  model  is trained with a training dataset.
Then, to test the performance of  
the method on predicting cases to which the model has not been exposed, a test dataset is used. The MSE for predicted $y$ (shown with $y^{\prime}$) in a test dataset, made of $n$ data points, is calculated as
\begin{equation}\label{eq:MSE}
    {\mathrm {MSE}}=\frac{\sum_i (y_i - y^{\prime}_i)^2}{n}.
\end{equation}
This is the mean of the squared residuals (i.e., the difference between the real and predicted values). When the model’s predictions are close to the real values, the MSE becomes smaller. To evaluate the performance of the model, the $R^2$ score is calculated as 
\small
\begin{equation}\label{eq:R2}
\begin{split}
    & R^2 = 1-\frac{\mathrm {RSS}}{\mathrm {TSS}},\\  \mathrm{\,\,where\,\,} \quad
    & \mathrm{RSS}  = \displaystyle \sum_i (y_i - y^{\prime}_i)^2,\\    \mathrm{\,\,and\,\,} \quad
     & \mathrm{TSS} = \displaystyle \sum_i \left( y_i - \frac{\sum_i y_i}{n} \right)^2.
\end{split}
\end{equation}
\normalsize
RSS is the sum of the squared residuals. 
%\sout{and} 
TSS is the total sum of squared residuals when the effects of other variables are not considered, 
%\sout{in which}
%\textcolor{red}
{so that} all the $y_i$  values are predicted to be equal to their mean ($y^{\prime}_i=\sum_i y_i/n$). The $R^2$ score measures the residual decrease when a model is employed compared to the simplest case of $y^{\prime}_i= \mathrm{mean}(y_i)$. $R^2$ is meaningful when $0 \leq R^2 \leq 1$. Models with predictions closer to the real values of dependent variables have higher $R^2$ scores. 

To be able to perform comparisons, we renormalize and non-dimensionalize the parameter values to the range [0, 1]. 
For example, feature X in the range $[X_\mathrm{min}, X_\mathrm{max}]$ is scaled using  
\begin{equation}\label{eq:standardize}
    X_{\mathrm {standard}} = \frac{X-X_\mathrm{min}}{X_\mathrm{max}-X_\mathrm{min}}.
\end{equation}
After evaluating model performances, predictions can be scaled back to their original range.

%%%%%%%%%%%%%%%%%%%%%%%%%%%%%%%%%%%
% \section{Machine Learning Methodology and evaluation
\subsection{ML model evaluation
\label{sec:ML_modelEvaluation}}

For ML, the data was split into two parts: 
1) to train the model using the train-dataset (viz.~80\% of the experimental data), and 
2) to test the model performance with the test-dataset (viz.~the remaining 20\%).
When the number of data points is limited, results can depend on 
%a particular random 
the choice of the training set. Therefore, to compare the models’ performances, a $k$-fold cross validation (cv) technique~\cite{hastie2009elements} was used to assure that the test and train datasets are representative of the same population of data, and also to track possible over- or under-fitting. In this technique, the training set is divided into $k$-folds of which ($k-1$) folds are used for training the model and the remaining fold is used to validate the models’ predictions. The validation fold is chosen one-by-one until all the $k$ folds have been tested once and the reported cv-MSE is the average of the MSEs for $k$ different validation folds.

Table~\ref{tab:methods} shows the cv-MSE for different models.  The table also reports the calculated test-MSE based on predictions of the cv-trained model and the initial test-split that was set aside at the beginning. The results in Table~\ref{tab:methods} provide a comparison of the linear and polynomial regression models, along with XGB
algorithm through their cv-MSE and test-MSE. The difference between these two MSEs shows the tolerance of the model performance on the different sets of unseen data. 

Once a model's cv-MSE is considerably smaller than its test-MSE, the model has overfitted, as is the case for the 3$^{\mathrm {rd}}$ order polynomial regression. 
The model underfits if both cv-MSEs and test-MSE are high, as is the case for the 4$^{\mathrm {th}}$ order polynomial regression, which means the model has not learned the correlations between the data.

Another technique to evaluate robustness of a model is bootstrapping. In this case an algorithm is trained multiple times with multiple random train-test splits~\cite{franklin2005elements} and the average MSE and $R^2$ are reported.
Table~\ref{tab:methods} also includes the bootstrap MSE and $R^2$ score for the above different models and a NN algorithm as well as the calculation times per model training. 

\begin{table*}
    % \footnotesize
    \caption{Performance of the different ML algorithms for predicting the channel’s depth.}
    \centering
% \toprule
% \begin{tabular}[]{|p{3.9 cm}|p{3 cm}|p{3 cm}|p{3.5 cm}|p{3 cm}|}
\begin{tabular}{| c | c| c | c | c |}
\hline
 {\bf ML model} & 10-fold cv-MSE $\times 100$ & test-MSE $\times 100$ & Bootstrapping MSE$\times 100$,   $R^2$ & {\bf calc. time}\,(s) \\
  \hline   \hline
 Linear Regression~(LR) & 2.007& 1.982 & 1.925, 67.60\%& 0.003\\ 
 \hline
 2$^{\mathrm{nd}}$ order Poly.R.& 1.136& 1.479& 1.111, 81.59\% & 0.004\\
\hline
 3$^{\mathrm{rd}}$ order Poly.R. & 0.676 & 1.769 & 1.096, 80.80\% & 0.009\\
 \hline
4$^{\mathrm{th}}$ order Poly.R. & 3.624 & 3.511 & 5.078 , 21.02\% & 0.022 \\
 \hline
 XGBoosting & 1.559 & 1.164 & 1.418, 75.74\% & 0.084\\
 \hline
Neural Networks & - & - & 0.687, 87.44 \% & 2.213\\
\hline     
\end{tabular}\\
\label{tab:methods}
\end{table*}

According to Table~\ref{tab:methods}, NN shows the lowest bootstrapping MSE and the highest $R^2$ score. It is followed by the 2$^{\mathrm {nd}}$ and 3$^{\mathrm {rd}}$ order polynomial regression models. The 3$^{\mathrm {rd}}$ order polynomial regression, however, is prone to overfitting as discussed above. Although NN requires a longer computation time compared to the other methods, the time is short ($\approx 2$~seconds). 
This supports the assertion in Ref.~\cite{mills2021lasers} that
%, in reported literature, 
NN predictions are to be preferred owing to their higher accuracy compared to other ML methods' for laser machining.

%%%%%%%%%%%%%%%%%%%%%%%%%
\subsection{Comparison of prediction methods}\label{sec:model_comparison}

\begin{figure*}
    \centering
    \includegraphics[width= 0.325\textwidth]{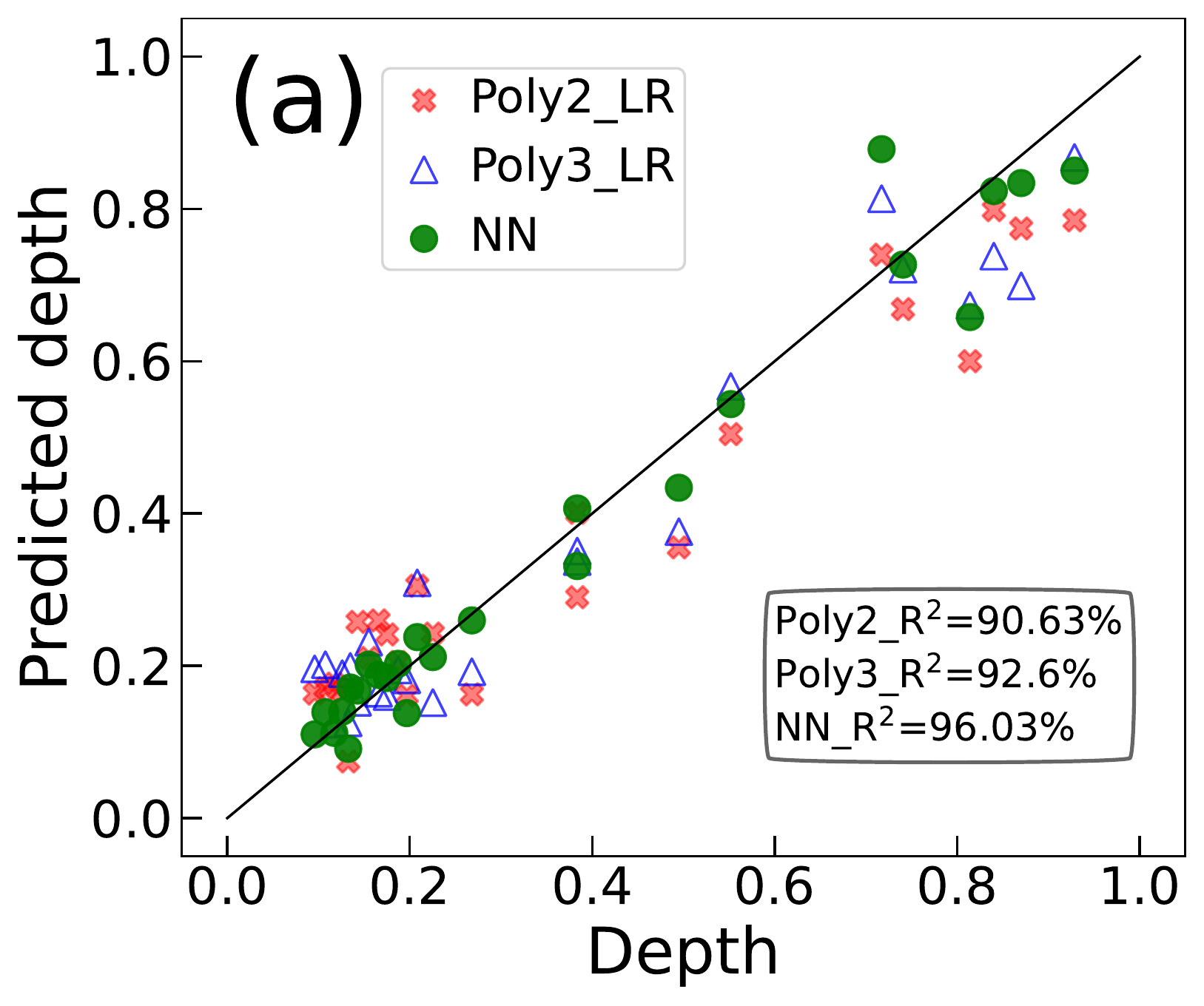}
    \includegraphics[width= 0.325\textwidth]{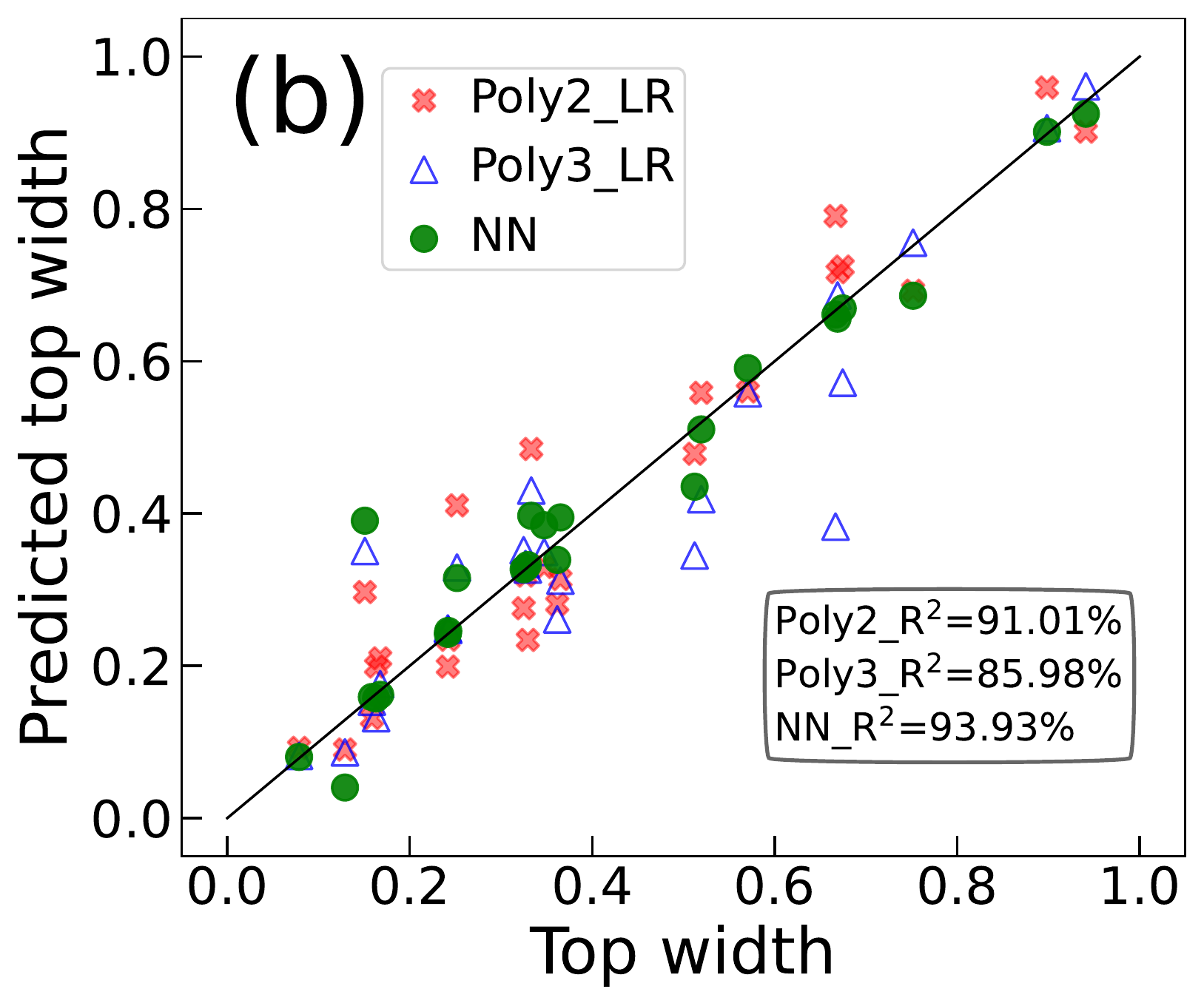}
    \includegraphics[width= 0.325\textwidth]{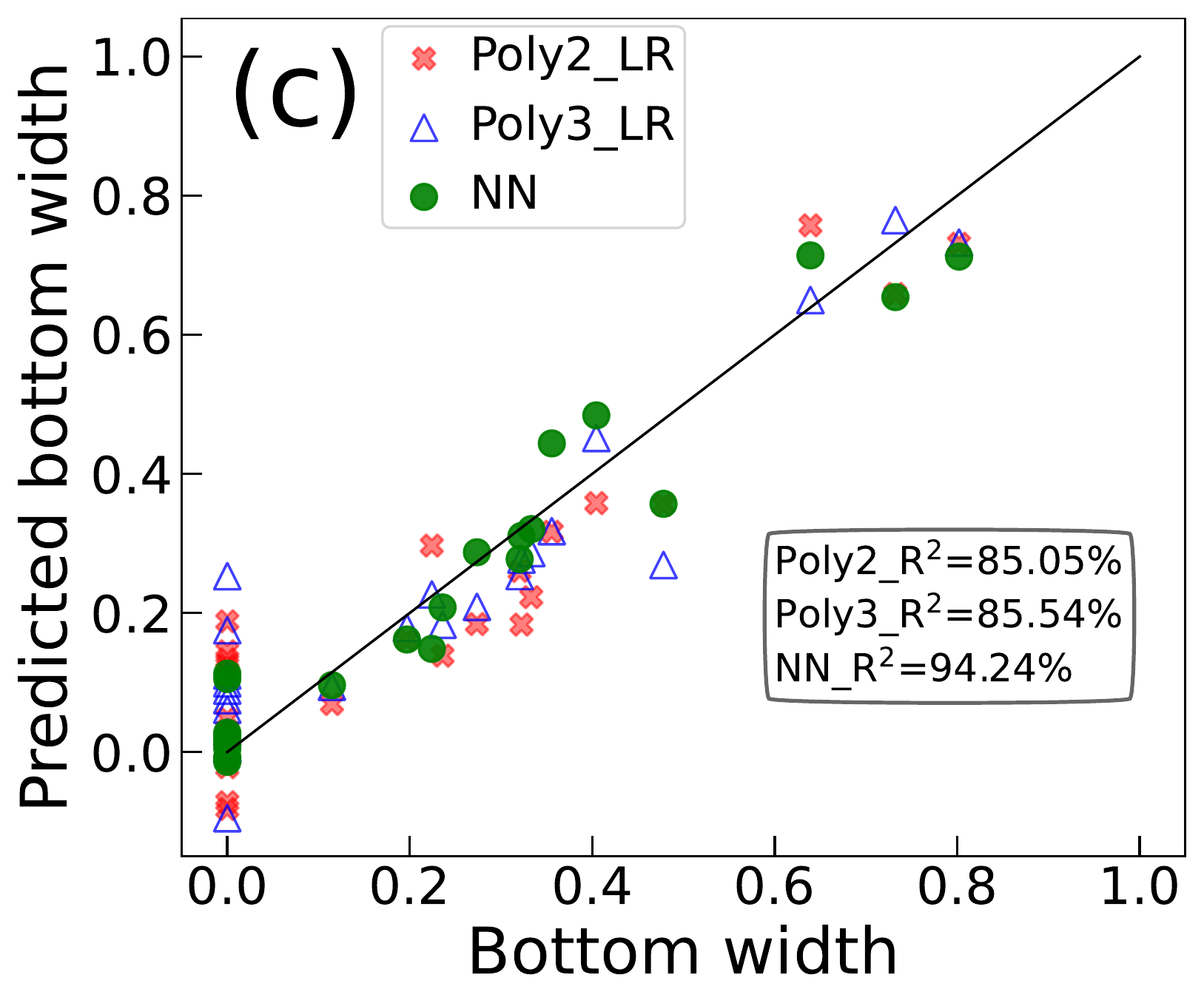}
    \caption{Comparison of the measured and predicted values for the channel (a) depth, (b) top width, and (c) bottom width with the $2^{\mathrm {nd}}$ and $3^{\mathrm{rd}}$ order polynomial regression models and neural network (NN). All of the input and output parameters were renormalized to [0,1] interval (see Eq.~\ref{eq:standardize}). Models were trained with 80\% of the experimental observations and tested with the remaining 20\%. The diagonal line is the limit of perfect prediction.}
    \label{fig:method_comp}
\end{figure*}

After comparing the general performance of the different ML methods using the experimental data, the best three methods (i.e., the NN, 2$^{\mathrm {nd}}$ order and 3$^{\mathrm {rd}}$ order polynomial regressions) were trained with 80\% of data and used to predict the channel dimensions of the test dataset (the remaining 20\%). The results are presented in Figure~\ref{fig:method_comp}. 
The predicted values of the different models are distinguished with different colours and symbols, and the diagonal black lines show the perfect prediction limit, when the predicted values match the real dimensions. The data in Figure~\ref{fig:method_comp} show that although the performance of the regression model for the widths is remarkable, the three models are at their best when predicting the depth (R$^2>90$\%). This can be explained by noting the fact that the 2$^{\mathrm {nd}}$ and the 3$^{\mathrm {rd}}$ order polynomial regression methods are modifying 17 (4 for parameters' first power, 4 for second power, 8 for interactions and one for intercept) and 21 (viz.~2$^{\mathrm {nd}}$ order's plus 4 for the third powers) 
coefficients, respectively, 
%\textcolor{red}
{as explained in Equations~\ref{eq:poly_regression},} to minimize the loss (viz.~MSE here 
%\textcolor{red}
{as in Equation~\ref{eq:MSE}}), but the selected NN structure modifies at least $4\times64+64\times32+32\times3=2400$ coefficients, which means more degrees of freedom to capture the correlations between the input laser parameters.

%%%%%%%%%%%%%%
\subsection{Feature importance}\label{sec:feature_importance}

%\sout{Various models interpret the importance of input parameters (features) for more accurate output predictions differently. One of the quantities the XGB algorithm provides is feature importance}

%\textcolor{red}
{
%Among different ML models, 
The XGB algorithm provides a quantity called feature importance} which is a score indicating how valuable each feature is in the construction of the boosted tree~\cite{hastie2009elements}. The score is the average of the IG scores attributed to each feature-based split over all the decision trees within the XGB model. 
%\textcolor{red}
{A feature has a higher importance score when the splits based on its values lead to more accurate predictions. }       

\begin{figure*}
    \centering
    \includegraphics[width= 0.325\textwidth]{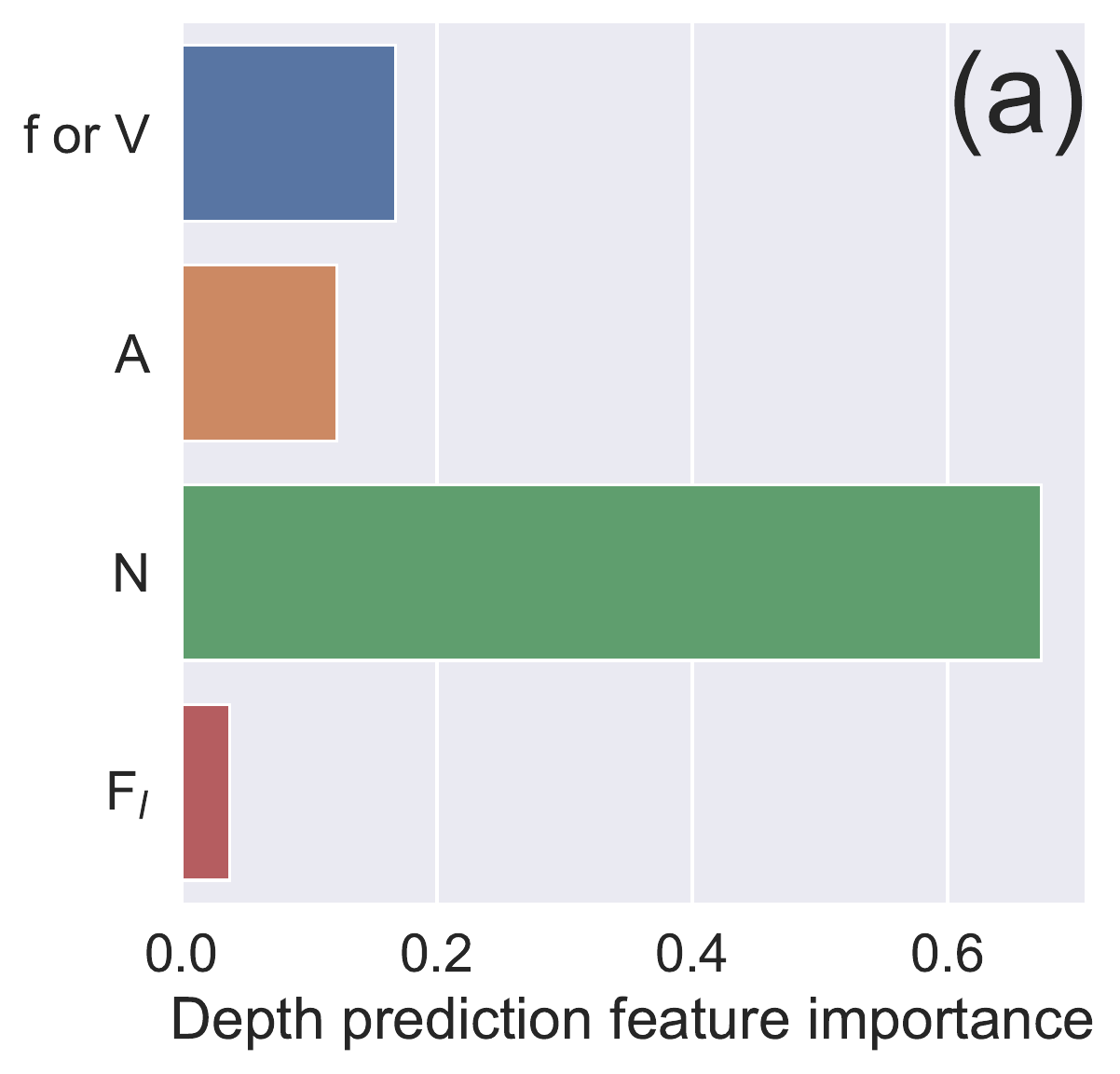}
    \includegraphics[width= 0.325\textwidth]{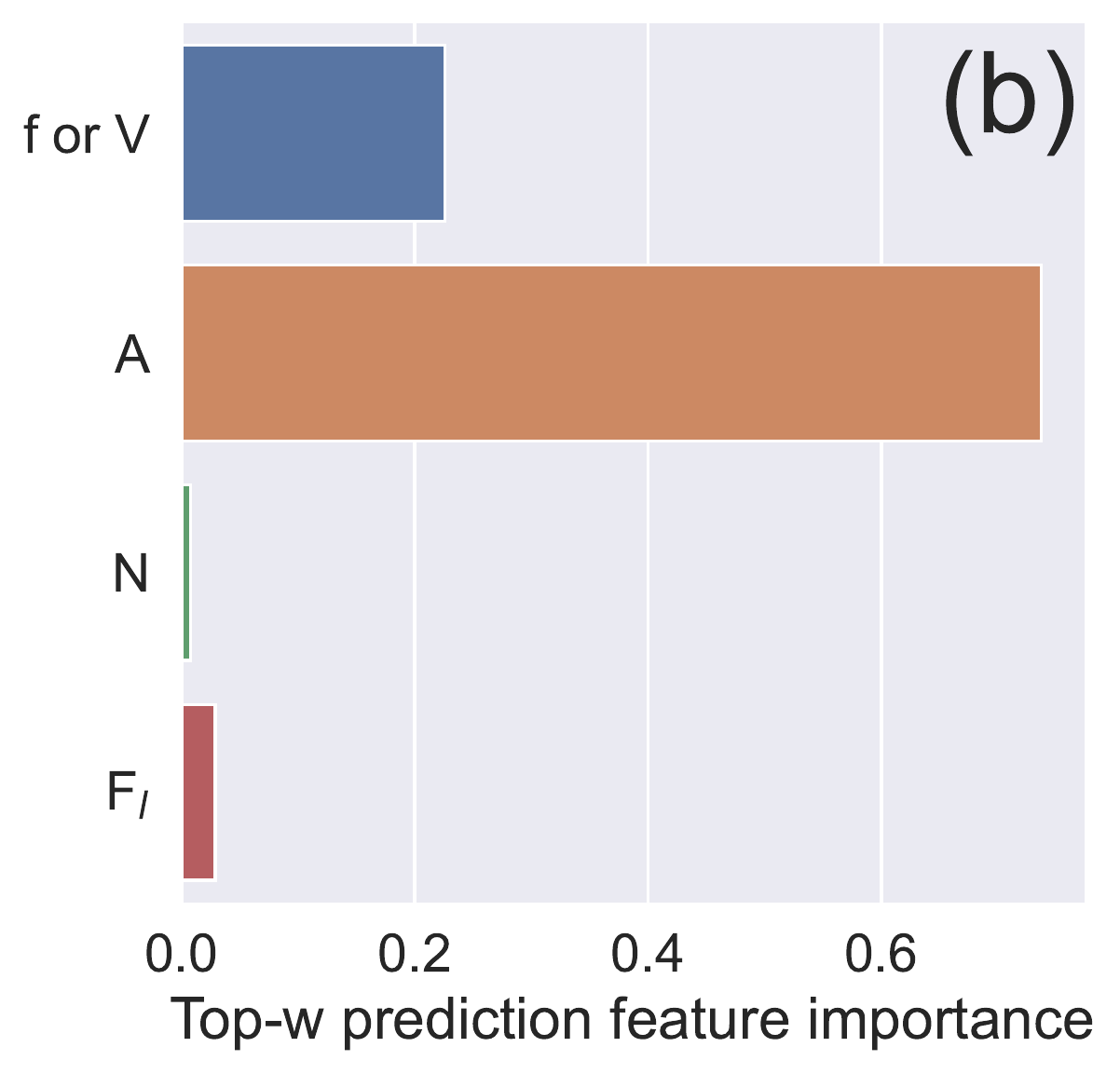}
    \includegraphics[width= 0.325\textwidth]{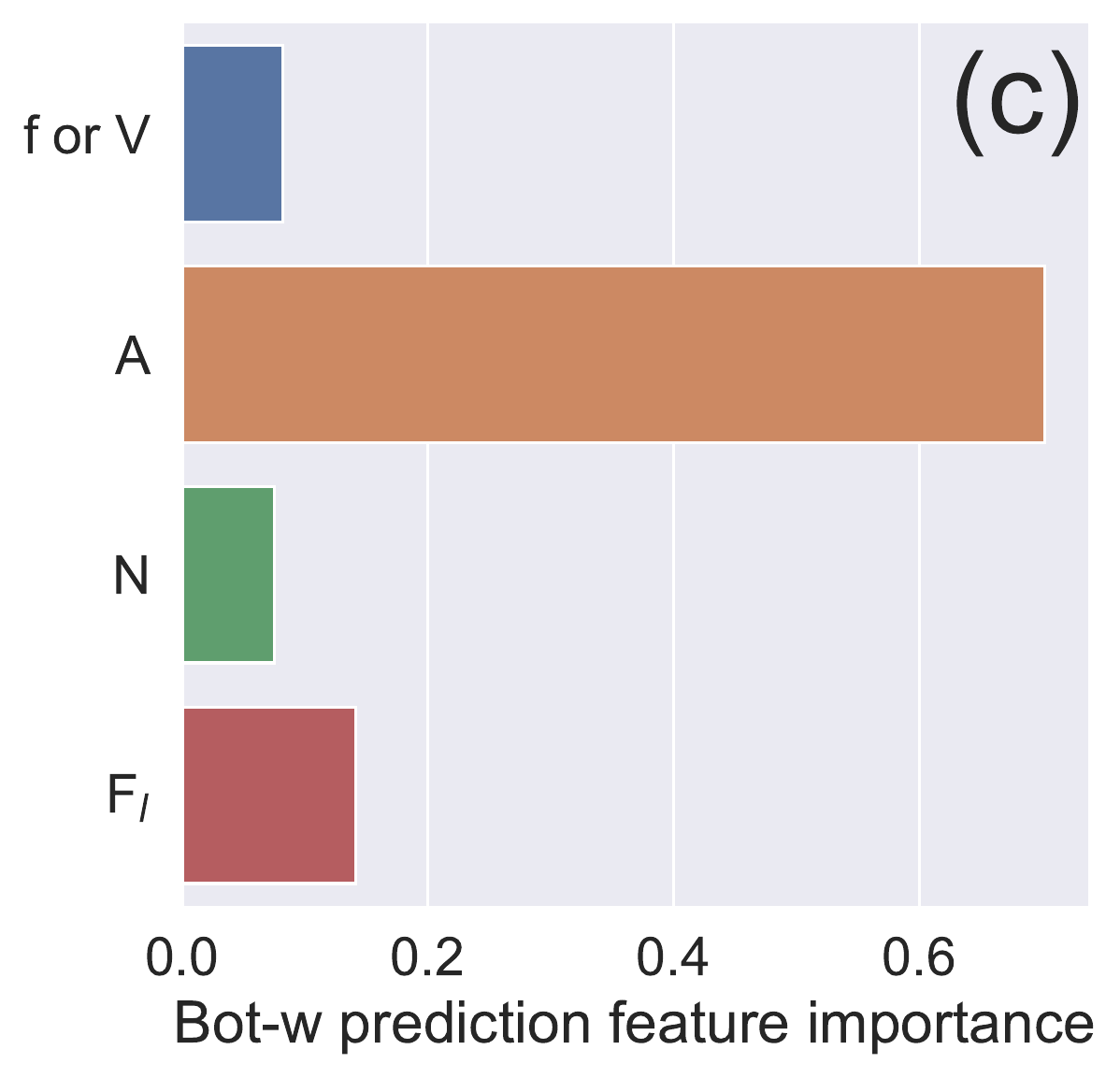}
    \caption{Variable importance in predicting the channel’s depth and widths by the XGB method~\cite{chen2016xgboost}. The most important factor for predicting the channel depth is the number of passes (N), while  for the channel width it is the laser beam amplitude (A).}
    \label{fig:XGB}
\end{figure*}

Although the XGB is not the highest performance model for this data, it is worth looking at the feature importance hierarchy it prescribes for predicting channel dimensions. 
Figure~\ref{fig:XGB} shows the XGB-determined feature importance for predicting the channel’s depth, top and bottom widths with respect to the laser parameter inputs. 

The variable importance identified by XGB can be understood from Figures~\ref{fig:amplitude} and~\ref{fig:passes}. As shown in Figure~\ref{fig:passes}, the channel depth sharply increases when the number of passes exceeds 40. This is consistent with the XGB result on assigning the number of passes ($N$) as the most important parameter for depth prediction, as shown in Figure~\ref{fig:XGB}(a). Similarly, Figure~\ref{fig:amplitude} shows that the behaviour of the channel’s top width changes when the amplitude is around 0.5\,mm. This supports the XGB assertion that data split based on amplitude ($A$) improves the  width predictions the most, see Figures~\ref{fig:XGB}(b) and (c). Preliminary experimental investigations show that more efficient laser machining is achieved via maintaining a constant ratio of the frequency to the linear speed. This is the reason why, this ratio has been kept fixed for all the experimental data. 
As a result, between frequency and linear speed, only one is considered independent.

%%%%%%%%%%%%%%
\subsection{NN structure}\label{sec:NN_structure}

\begin{figure}
    \centering
     \includegraphics[width= 0.495\textwidth]{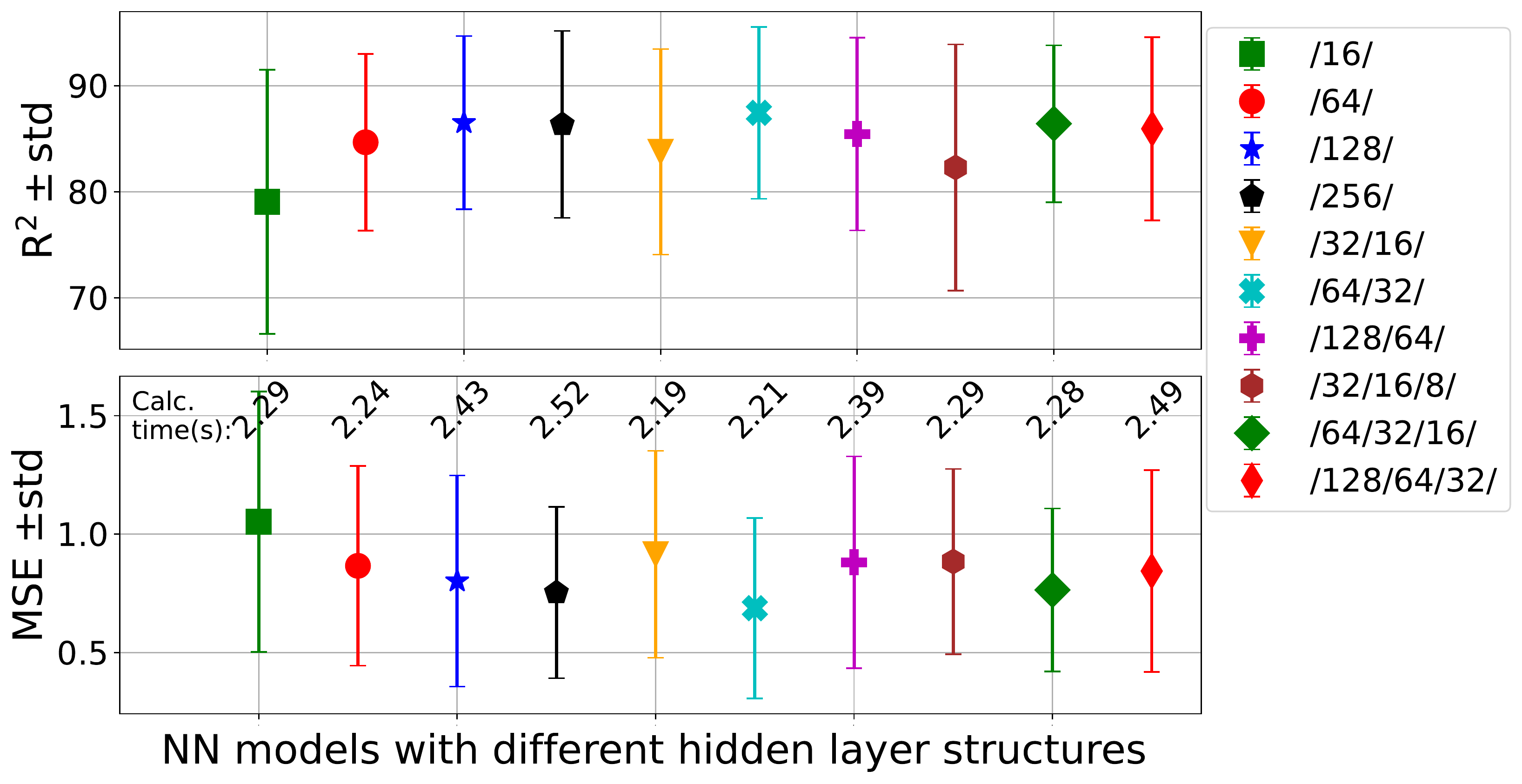}
    \caption{Comparing 10 different structures of NN by changing the number of hidden layers and neurons per layer (as printed in the legends) for predicting the channel’s depth using four independent laser parameters. Owing to the limited number of datapoints, the NN predictions’ performances depend on the training and test datasets. To avoid this problem 100 independent 80-20 splits (bootstrapping) were run and the mean values of $R^2$ and MSE are shown within $\pm$std deviation as error bars. The calculation time per run for each structure is printed on the bottom panel above their MSE error bars.}
    \label{fig:NNstructure}
\end{figure}
%\textcolor{red}
{In order to find the most efficient NN structure for this problem, 10 different structures were examined}
by varying the number of hidden layers and neurons per layer, as indicated in the label of Figure~\ref{fig:NNstructure}. Owing to the limited number of observations, the bootstrapping technique was used, 
and the mean 
%\textcolor{red}
{values of} $R^2$ scores and MSEs 
%\textcolor{red}
{(based on Equations~\ref{eq:MSE} and \ref{eq:R2})} are reported. The error bars indicate the 
standard deviation interval for each 
%\textcolor{red}
{arrangement. The average calculation time for each case, printed above its MSE error bar, shows roughly similar values for all the tried structures}. 
Increasing the number of hidden layers or neurons
%\textcolor{red}
{per layer} provides more degrees of freedom, which can result in more accurate predictions, but beyond a certain point, the effect diminishes. As shown in Figure~\ref{fig:NNstructure}, the mean values of the $R^2$ scores and MSEs for depth predictions as well as the calculation time
are not significantly different for the structures investigated here. 
As a result,
the hidden layer structure of /64/32/ with higher $R^2$ score and smaller MSE is used for the rest of NN predictions.

%%%%%%%%%%%%%%%%%%%%%%%%%%%%%%%%%%
\section{Performance evaluation and application}\label{sec:application}

Since NN proved to be the most successful approach for predicting the channel dimensions, 20 new laser combinations, in the same range as previous experiments, were used to compare the predictions of the best performing NN algorithm with the experimental values. The consistency of the experiments was evaluated by repeating the laser machining tests with fixed laser parameters to find the variations in dimensions. 
%\textcolor{red}
{Given the distribution of the measured dimensions, }the experimental uncertainty was found to be about 10\%.

\begin{figure*}
    \centering
    \includegraphics[width= 0.9\textwidth]{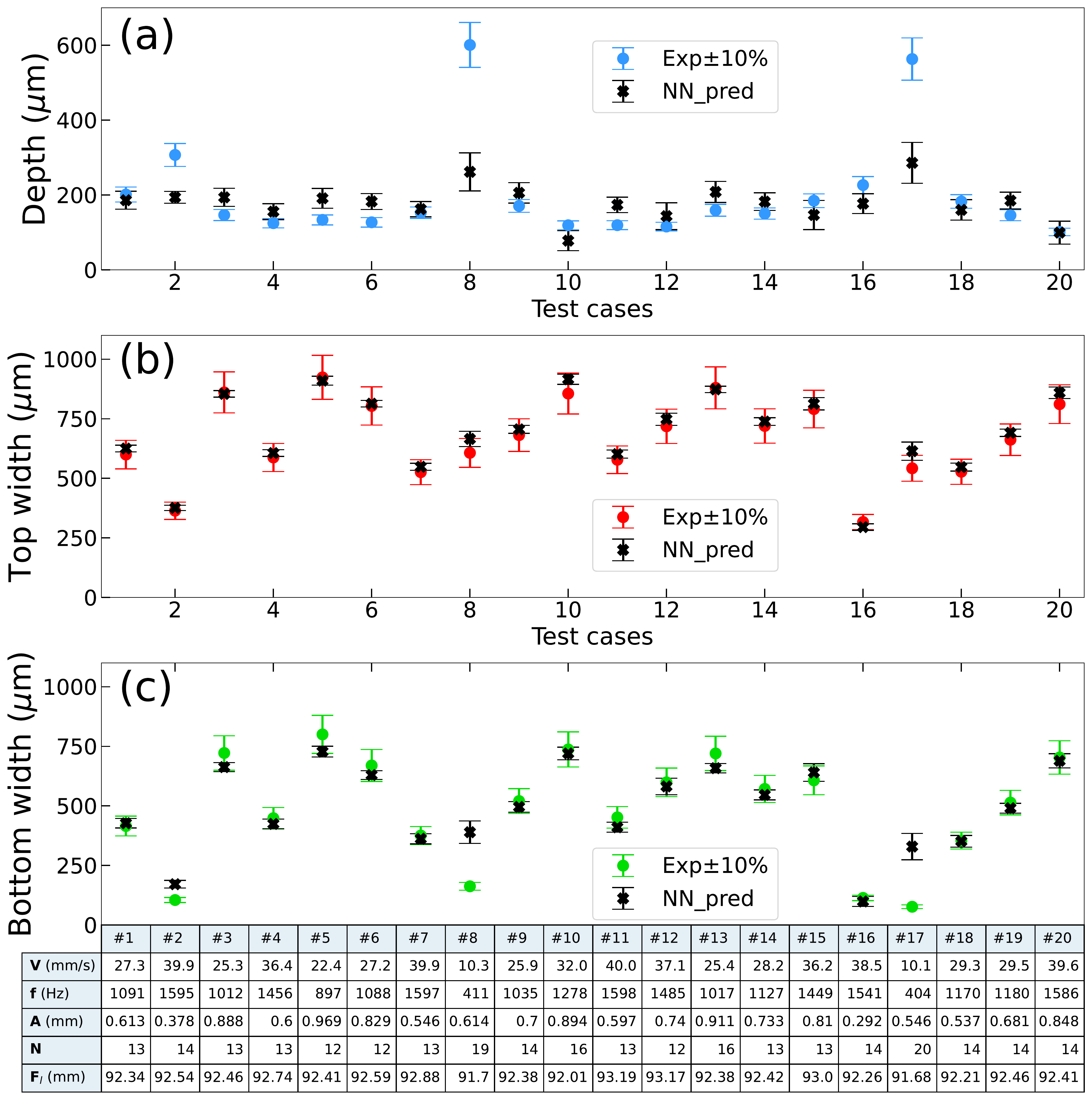}
    \caption{Comparison of the predicted values of channel dimensions to the experimental results for 20 laser combinations. Experimental results are shown with 10\% uncertainty and NN predictions are presented as the mean value of 100 different initialization with one standard deviations above and below, represented as error bars. The laser parameters for test cases are printed in the table.}
    \label{fig:test_preds}
\end{figure*}

\begin{figure*}
    \centering
        \includegraphics[width= 0.45\textwidth]{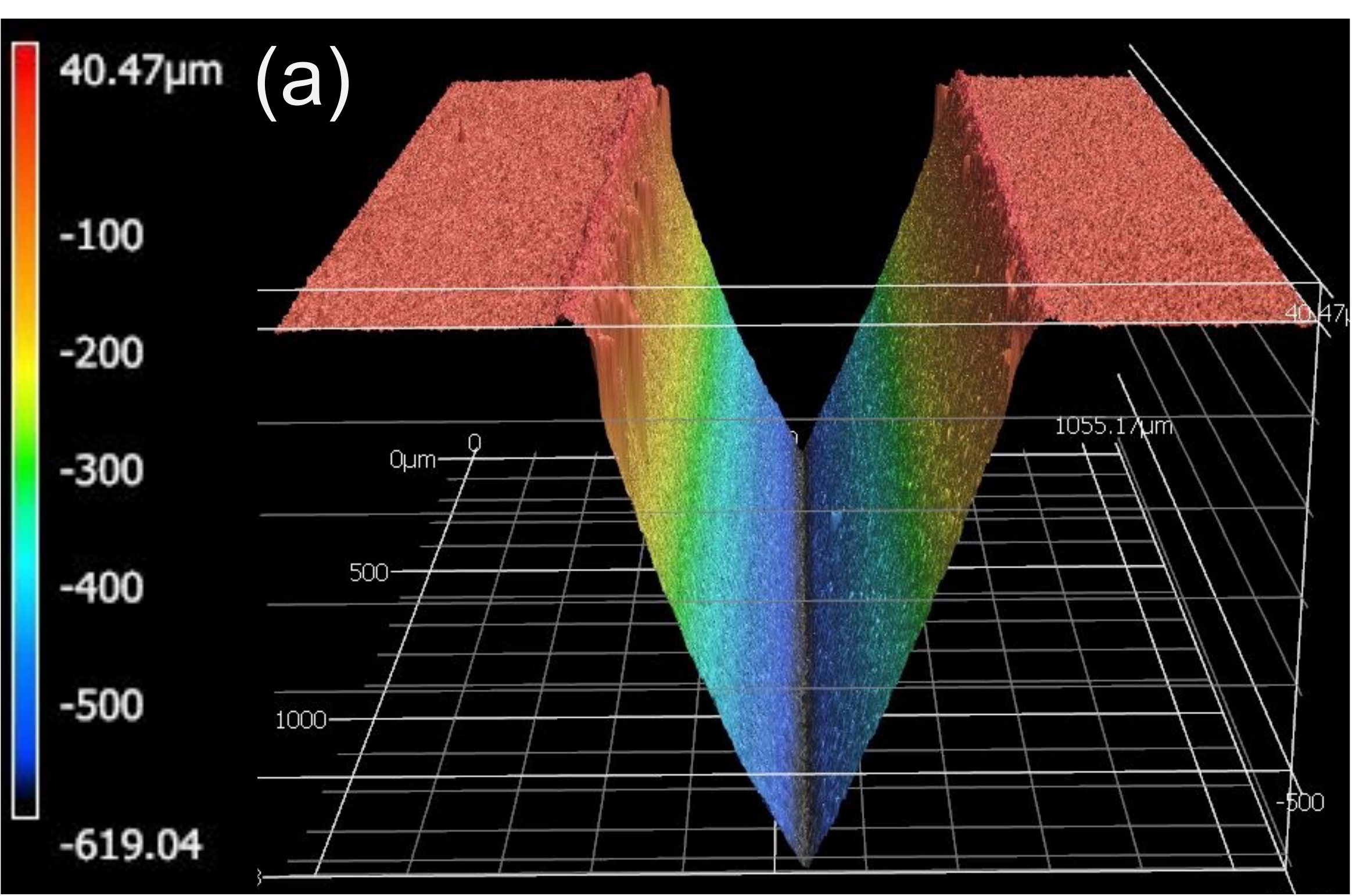}
    \includegraphics[width= 0.45\textwidth]{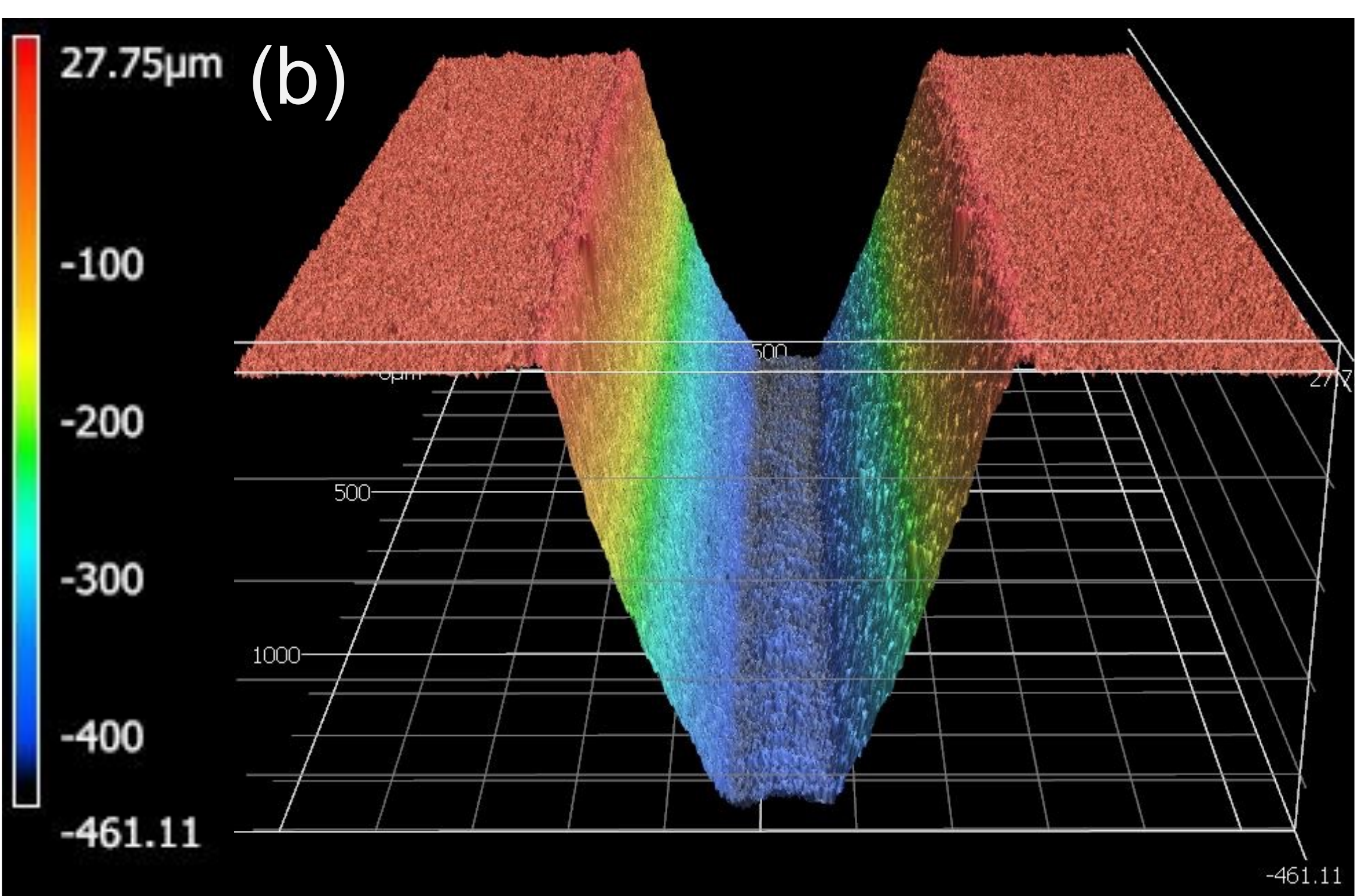}
            \includegraphics[width= 0.45\textwidth]{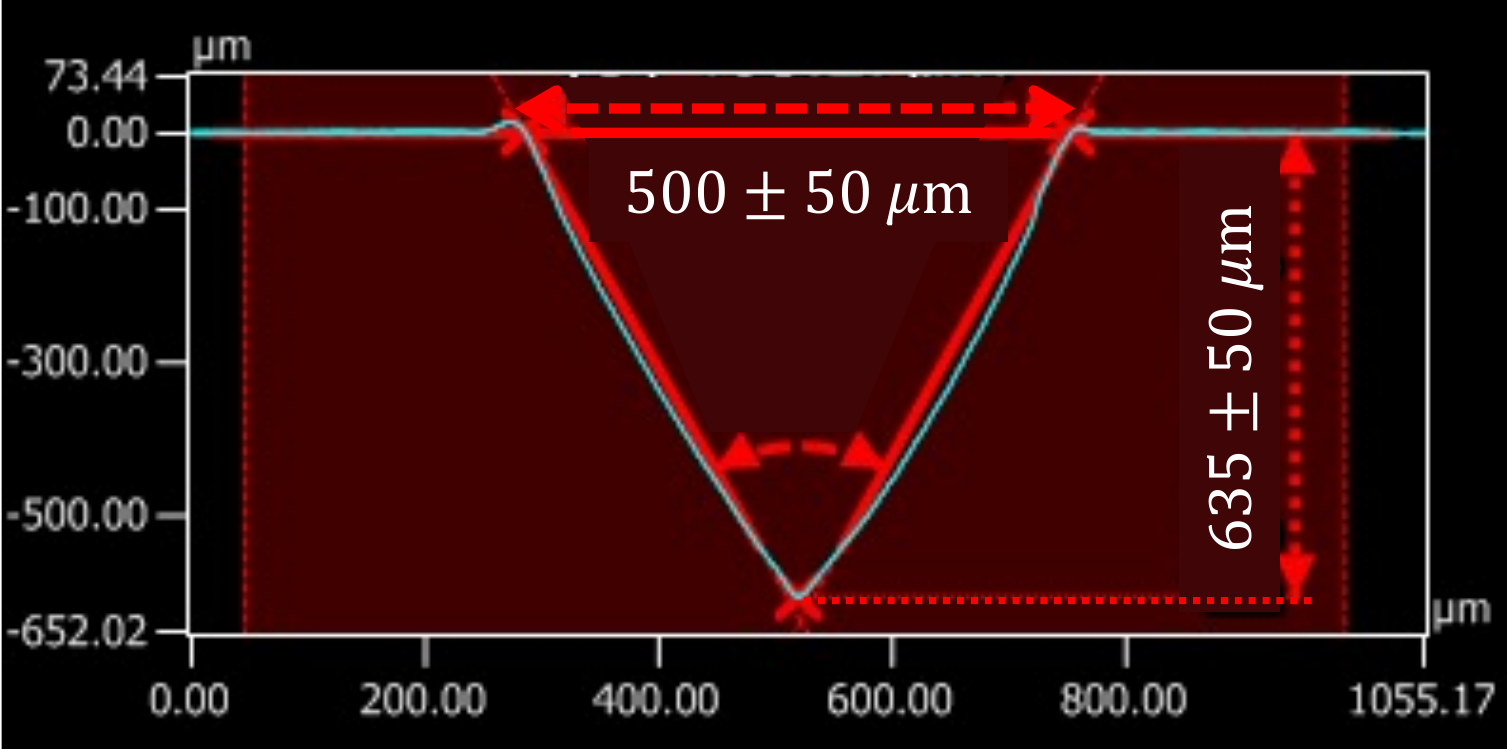}
    \includegraphics[width= 0.45\textwidth]{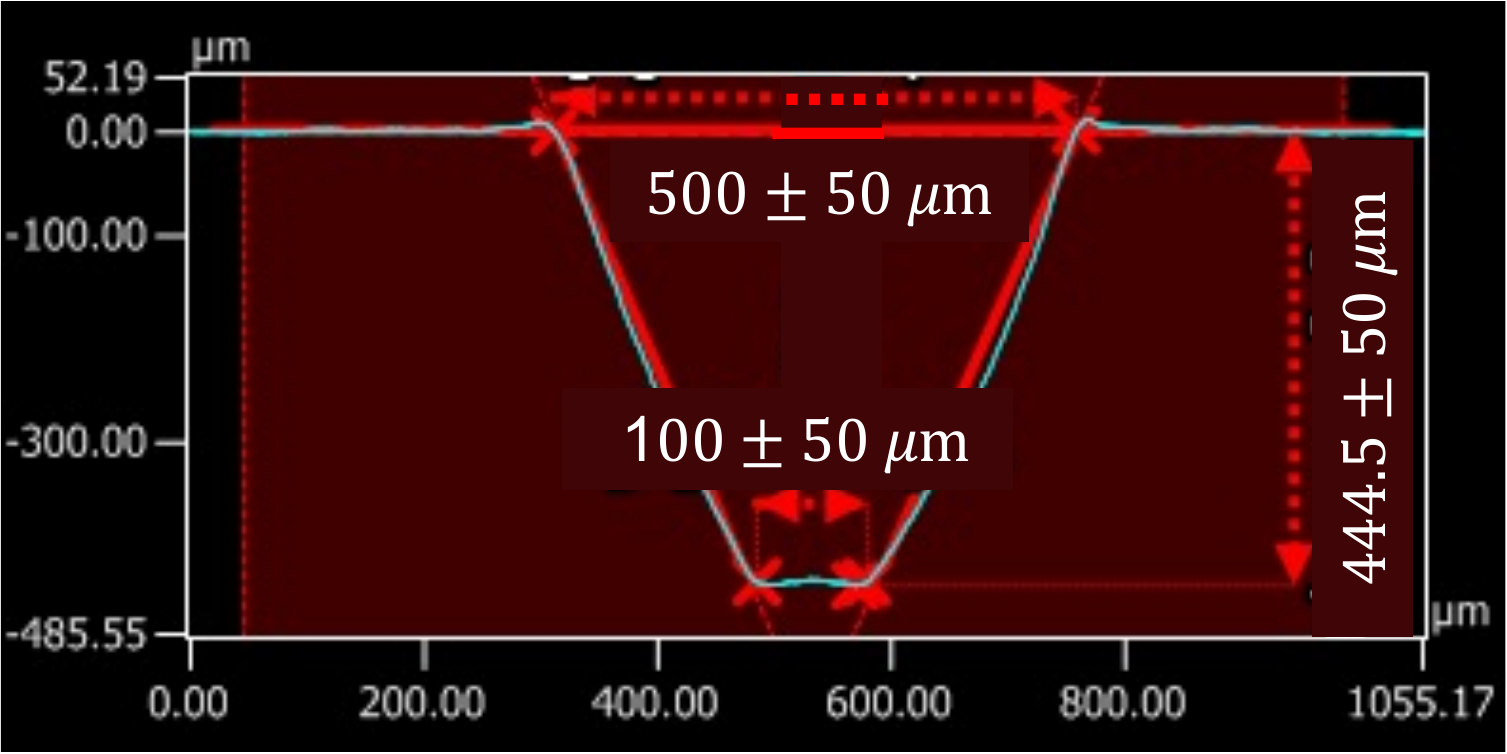}
    \caption{Three and two dimensional representations of typical engraved channels with target dimensions printed having (a) triangular and (b) trapezoidal cross-sections.}
    \label{fig:3D_2Dcuts}
\end{figure*}

\begin{figure}
    \centering
    \includegraphics[width= 0.45\textwidth]{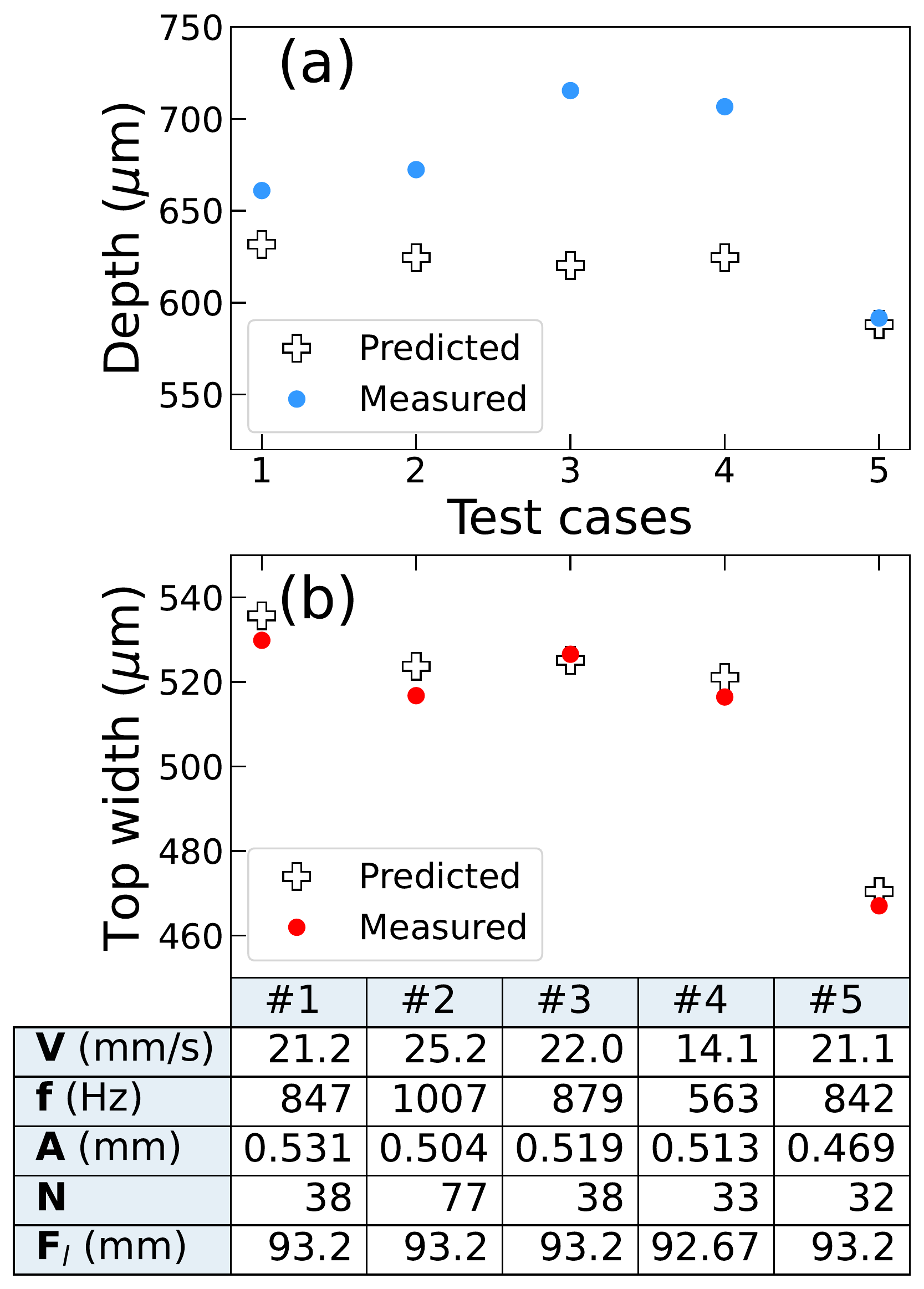}
    \caption{Model predictions for a triangular cut with depth $635\pm50~\mu$m and top width $500 \pm 50 \mu$m. The laser parameters for each case is printed in the Table.}
    \label{fig:TargetTriangle}
\end{figure}

\begin{figure}
    \centering
    \includegraphics[width= 0.5\textwidth]{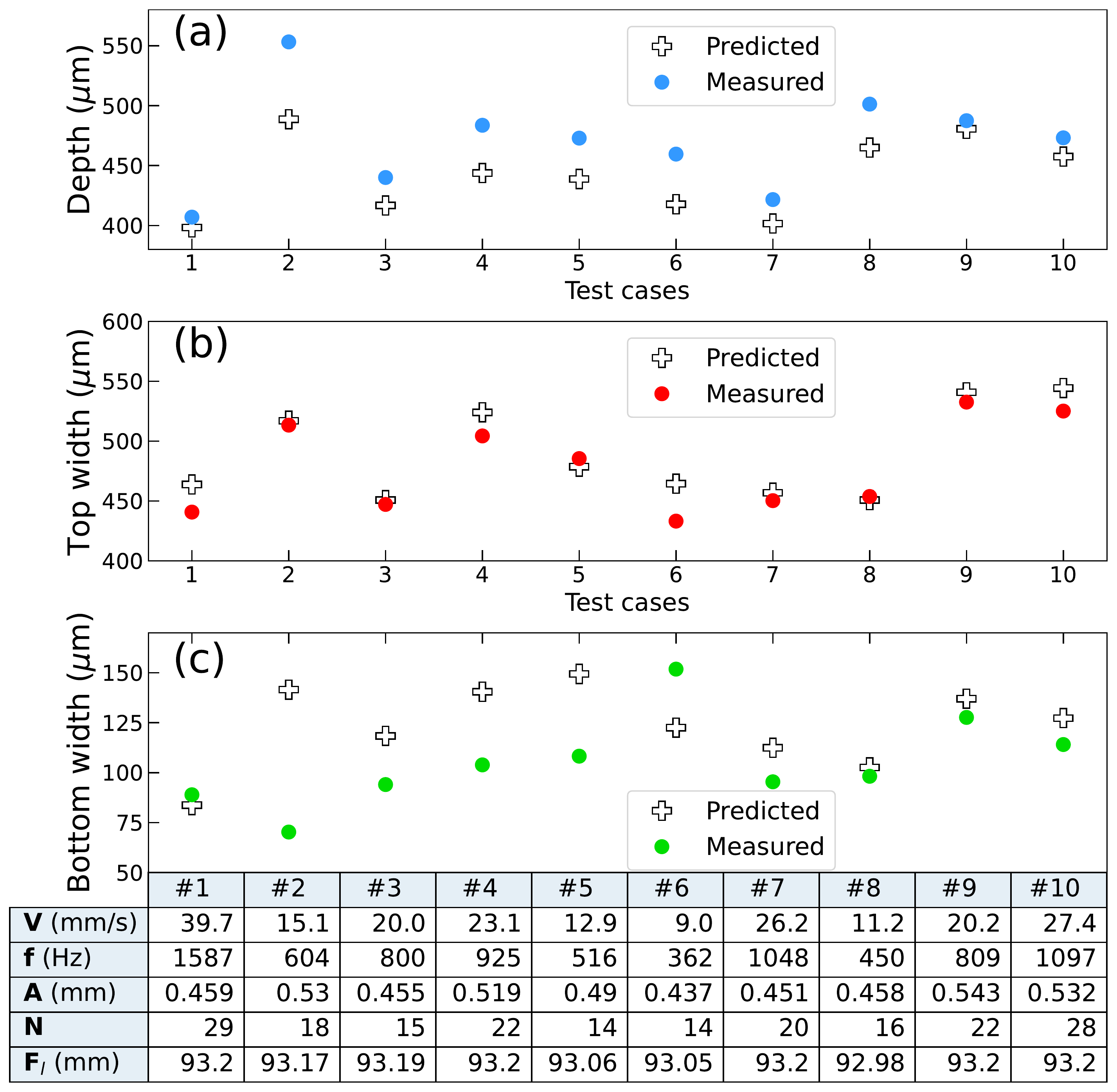}
    \caption{Model predictions for a trapezoidal cut with depth $444.5\pm50~\mu$m, top width $500 \pm 50 \mu$m and bottom width $100 \pm 50 \mu$m. The laser parameters for each case is printed in the table.}
    \label{fig:TargetTrap}
\end{figure}
NN predictions depend on the neurons’ initial weights, which are being modified through back propagation. This dependence was negated by running the NN for 100 random initializations. The mean of these 100 predictions is shown
within error bars (standard deviation) of the predictions
in Figure~\ref{fig:test_preds}.
%\textcolor{red}
{This figure also indicates that in 17 out of 20 test cases at least two out of three measured dimensions overlap the NN prediction (except for case 2, 8 and 17).}
% The depth predictions, shown in Figure~\ref{fig:test_preds}(a), indicates that in 17 out of 20 test cases the experimental depth overlaps the NN prediction intervals (except for cases 2, 8 and 17). At the same time, as shown in Figure~\ref{fig:test_preds}(b), the measured top widths match the predicted values in all cases, while the measured bottom widths for cases 2, 8 and 17 do not overlap the NN prediction intervals, as shown in Figure~\ref{fig:test_preds}(c). 
Comparing the laser parameters for the different test cases, one can see that the employed frequency for the cases with the largest variations from the predicted values, case number 8 and 17, is around 400~Hz, whereas for the rest of the cases it is in the range of $[897, 1600]$~Hz. Moreover, the laser-substrate distance for the cases 8 and 17 is roughly 91.7~$\mu$m while for the rest it is between 92 to 93.2~$\mu$m. 

Detecting correlations between laser parameters and reliably predicting the channel dimensions enables the designer to find the proper input parameters for the desired channel with ease. In order to achieve this, one needs to make a collection of applicable laser parameters and feed it to an ML model that is trained with all the available experimental observations. 
By filtering the target dimensions among the model predictions, one can capture the parameter combinations that lead to target outputs. Laser parameter combinations can be made as a fine grid sweeping the range of the parameter values used in experiments~\cite{mcdonnell2021machine,teixidor2015modeling, dhupal2007optimization}. 

Parameter combinations are applicable to the laser machine when the restrictions on the input parameters are known. 
%\textcolor{red}
{However, when}
the constraints are unknown, generating synthesized laser combinations is an alternative.
Different techniques of data generation are categorized into the older methods such as synthetic minority over-sampling technique (SMOTE)~\cite{chawla2002smote} and variational autoencoder~\cite{kingma2019introduction} focusing on balancing the data distribution close to the minority classes, and the recent innovation of generative adversarial networks (GANs)~\cite{goodfellow2014generative} which is used here owing to the high quality of the synthesized data. In this technique, two NNs are coupled and trained with real experimental observations to generate laser combinations similar to the training data~\cite{creswell2018generative}. Based on the type of training data, the standard GAN algorithm can be adjusted to improve the convergence problems~\cite{fekri2020generating}. 

To find the laser parameters for target channels, rather than a grid of 54,000 combinations, 5,000 parameter combinations were generated using the GAN algorithm and fed to the trained NN. Two target channels with triangular and trapezoidal cross-sections were specified; their 3D profiles and 2D scans are shown in Figure~\ref{fig:3D_2Dcuts}. The required laser parameters corresponding to these target channels were found by filtering back the NN predictions on the GAN data. To engrave the first target channel, having triangular cross-section with a depth of $635\pm50~\mu$m and top width of $500\pm50~\mu$m (see Figure~\ref{fig:3D_2Dcuts}(a)), five different laser combinations were tested as shown in Figure~\ref{fig:TargetTriangle}.
Given the limited number of training data and the diversity of the prescribed laser combinations, the agreement between the predicted and measured dimensions for different test cases is impressive.

Similarly, Figure~\ref{fig:TargetTrap} shows the measured and predicted dimensions for ten different laser combinations, printed in the enclosed table, corresponding to the second target channel with a depth of $444.5\pm50~\mu$m, top width of $500\pm50~\mu$m and bottom width of $100\pm50~\mu$m (see Figure~\ref{fig:3D_2Dcuts}(b)). Same as the first target set, disparity of the prescribed laser combinations and their significant agreement proves the capability and efficiency of the used NN algorithm for ceramic machining applications.

%%%%%%%%%%%%%%%%%%%%%%%%%%%%%%%%%%%
\section{Conclusion}\label{sec:conclusion}

{In the present study we report a comparison and  application of ML methods to capture the interconnections and dependencies between the picosecond laser parameters for engraving channels on alumina ceramics. Systematic brute force laser parameter exploration 
%would otherwise not be 
is not feasible due to costly experiments and time-consuming simulations. 
}

%\textcolor{red}
{
Here, ML models were applied to the 
%interconnected and 
tunable parameters of a laser system. The parameters include the beam frequency and amplitude, the number of the laser passes over the substrate and the vertical distance of the light source from the substrate.
%to provide a better understanding of the output parameters such as the engraved channel’s depth, top width and bottom width.
}

%\textcolor{red}
{This study was performed in three stages. Stage one was the collection of experimental data on the produced channels created by ytterbium picosecond laser machining of industrial alumina ceramics. In stage  two the experimental data were used to train ML models for predicting 
%engraved 
channel dimensions.
%with an acceptable precision. 
For this purpose, the performances of different ML algorithms including linear/polynomial regression, XGB and NN, were compared. Although second or third order polynomial relationships between the input and output parameters are not excluded (considering the fairly high $R^{2}$ score of predictions via 2$^{\mathrm{nd}}$ and 3$^{\mathrm{rd}}$ order polynomial models), the complex relation suggested by the NN model provides a higher $R^{2}$ score and lower MSE. As the NN predictions depend on the coefficient initialization, the mean values of the predicted geometries of 20 test sets were exhibited to be within one standard deviation of NN predictions from 100 different initialization and demonstrated an overlap in most cases with their experimental uncertainty. The structure of the NN model (4/64/32/3) was justified by testing the performance of 10 different NN structures on the experimental data. The feature importance was studied via the XGB method which showed that the number of passes and amplitudes were the most determinative parameters for predicting the channels' depths and widths, respectively. These results are supported by experimental data in Section~\ref{sec:param_dep}. }

%\sout
%{The third and final stage was to make a collection of parameter sets applicable to the laser machine and feed it to the trained ML model to predict dimensions of the potential engraved channels. Then, one can filter out the input parameters corresponding to the target channel dimensions among the model predictions.  Owing to the unknown correlations between the parameters, not all the parameter sets are applicable to the laser. Therefore, generative adversarial networks (GAN) were used to collect parameter sets to apply to the trained model.}
%\textcolor{red}
{In the third stage, a collection of unseen laser parameters was fed to the trained model to make a table of required parameters for engraving target channels.   Generative adversarial networks (GAN) were used to collect the parameter sets, owing to the unknown correlations between the parameters.
By choosing a hand-full of parameter combinations corresponding to two individual target channels with triangular and trapezoidal cross-sections, the model's performance for general applications was examined.}           
Good agreement between the target geometries and the experimental cuts (based on predicted ML input parameters) demonstrated the success of the ML program for laser machining of ceramics.

\section*{Competing interests}

The authors declare no competing interests.

\section*{Author contributions}

All authors contributed to the design, implementation and analysis of the research, interpretation of the results, and writing of the manuscript. All authors have given their approval for the final
version of the manuscript.

\section*{Supplementary information}

In Figures~\ref{fig:amplitude}-\ref{fig:frequency} groups of laser combinations are distinguished with differently coloured symbols corresponding to sets of fixed laser parameters. These values are reported in Figures~S1-S4. 
%\textcolor{red}
{The experimental data collected and used in the ML analysis is reported in Tables~SI-SV.}

%%%%%%%%%%%%%%%%%%%%%%%%%%%%%%%%%%%%%
\section*{Acknowledgment}
The authors acknowledge E. Poirier, W. Amsellem, C. Beausoleil and Z. Katz for their technical assistance in the laser processing. RB  thanks M.N. Fekri for his guidance and useful discussions on ML results. The authors thank the New Beginnings Ideation fund at the National Research Council Canada (NRC). MK also acknowledges the financial support by the Natural Sciences and Engineering Research Council of Canada (NSERC) and Canada Research Chairs Program. 

%%%%%%%%%%%%%%%%%%%%%%%%%%%%%%%%%%%%
\bibliography{ms.bib}

%\printbibliography

\end{document}